\documentclass[12pt,preprint]{aastex}

\shorttitle{}
\shortauthors{Nesvorn\'y et al.}

\begin{document}
\baselineskip 19.pt

\title{Binary Survival in the Outer Solar System}

\author{David Nesvorn\'y$^1$, David Vokrouhlick\'y$^2$}
\affil{(1) Department of Space Studies, Southwest Research Institute,\\
1050 Walnut St., Suite 300, Boulder, CO, 80302, USA}
\affil{(2) Institute of Astronomy, Charles University,\\ 
V Hole\v{s}ovi\v{c}k\'ach 2, CZ--18000 Prague 8, Czech Republic}

\begin{abstract} 
As indicated by their special characteristics, the cold classical Kuiper belt objects (KBOs) 
formed and survived at $\simeq$42-47 au. Notably, they show a large fraction 
of equal-size binaries whose formation is probably related to the accretion of KBOs 
themselves. These binaries are uncommon in other --hot, resonant, scattered-- populations, 
which are thought to have been implanted from the massive disk below 30 au to $>$30 au during 
Neptune's migration. Here we highlight the possibility that equal-size binaries formed in the 
disk but were subsequently removed by impacts and/or dynamical effects (e.g., scattering 
encounters with Neptune). We determine the dependence of these processes on the size and 
separation of binary components. Our results indicate that tighter binaries, if they formed 
in the massive disk, have relatively good chances of survival (unless the disk was long-lived). 
In contrast, the widest binaries in the hot population, such as 2002 VF130, have a very low 
survival probability ($<$1\%) even if the massive disk was short-lived. They may represent a 
trace of lucky survivors of a much larger population of the original disk binaries, or they formed 
at $\sim$30-40~au and dodged the impact- and encounter-related perturbations that we studied here. 
We find that all known satellites of the largest KBOs would survive during the dynamical 
implantation of these bodies in the Kuiper belt. The low orbital eccentricities of Pluto's 
small moons may have been excited by impacts and/or encounters of the Pluto system 
to Neptune. 
\end{abstract}


\section{Introduction}

\begin{figure}
\epsscale{0.6}
\plotone{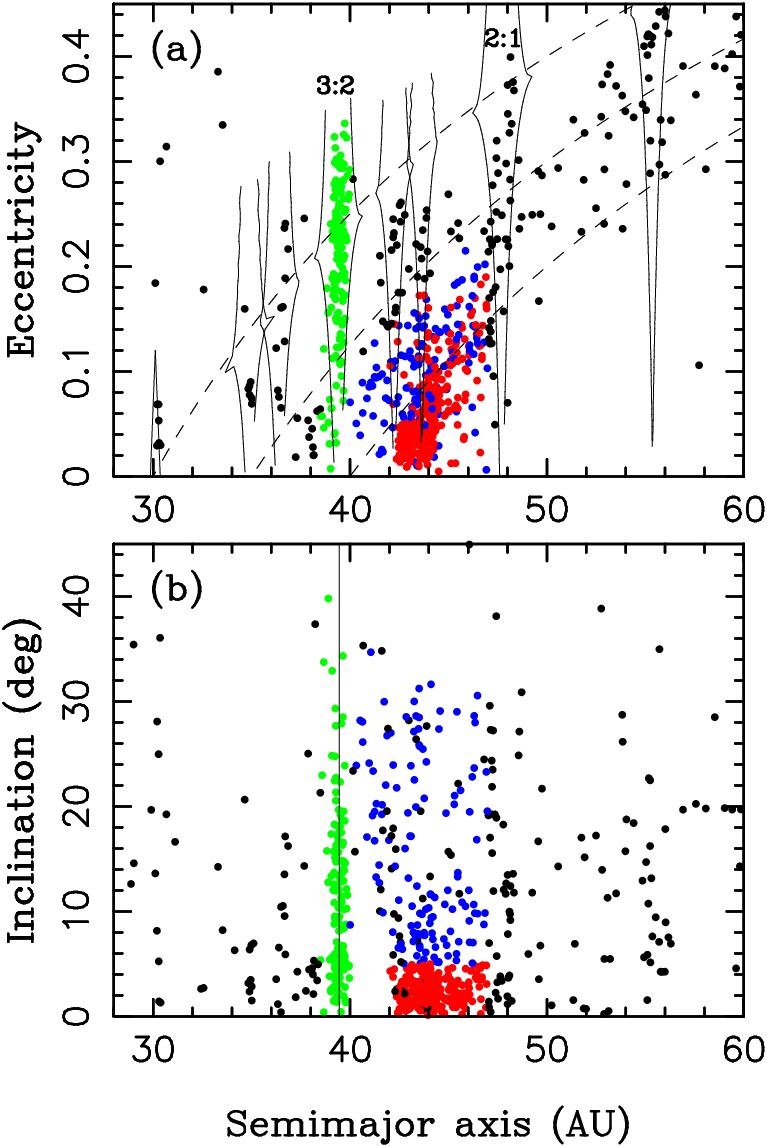}
\caption{The orbits of KBOs observed in three or more oppositions. Various dynamical classes are highlighted. HCs 
with $i>5^\circ$ are denoted by blue dots, and CCs with $i<5^\circ$ are denoted by red dots. The solid lines 
in panel (a) follow the borders of important orbital resonances. Plutinos in the 3:2 resonance are highlighted 
by green dots. The dashed lines in panel (a) correspond to $q=a(1-e)=30$, 35 and 40 au. See section 4.1 for 
definitions of different dynamical classes.}
\label{real}
\end{figure}

The Kuiper belt is a population of icy bodies beyond the orbit of Neptune (Figure~\ref{real}). 
The orbital structure of the Kuiper belt is an important constraint on the early evolution 
of the Solar System. It is thought that much of this structure, with large resonant populations 
and dynamically excited orbits, has emerged as a result of Neptune's migration into an outer disk 
of planetesimals (e.g., Hahn \& Malhotra 2005, Levison et al. 2008). The massive planetesimal 
disk that presumably existed below 30~au was completely dispersed by Neptune, and a small fraction 
of the scattered population was implanted onto orbits beyond 30~au, where it overlaps in 
orbital space with a population of bodies that formed and survived at $>$30~au (e.g., Batygin 
et al. 2011). Here we study an important tracer of this process, the KBO {\it binarity}, to set 
constraints on the initial binary fraction, planetesimal disk lifetime and timing of 
planetary migration.

Observations provide direct evidence for two different populations in the Kuiper belt. 
On one hand, Plutinos in the 3:2 resonance with Neptune, other resonant populations, Hot 
Classicals (HCs) and Scattered Disk Objects (SDOs; see Gladman et al. 2008 for a definition 
of these categories) share similar physical attributes. These populations (hereafter the dynamically 
{\it hot} KBOs) are thought to have been implanted into the Kuiper belt from below 30 au. 
The Cold Classicals (CCs), on the other hand, have low orbital inclinations (Brown 2001, Gulbis et al. 
2010), distinctly red colors (Tegler \& Romanishin 2000), relatively high albedo 
(Brucker et al. 2009, Vilenius et al. 2014), 
and a size distribution that shows a very steep slope at large sizes (Bernstein et al. 2004). 
The CCs are believed to have formed at $\simeq$42-47 au where they are found now.

Another important difference between the dynamically hot and cold populations in the Kuiper belt is the 
existence and nature of {\it binary} objects. A very large fraction of known 
100-km-class CCs are resolved binaries with nearly equal-size components (Noll et al. 2008a,b, 
Fraser et al. 2017, Grundy et al. 2018). These binaries are thought to have formed during 
the formation of KBOs themselves or by early capture (e.g., Goldreich et al. 2002,
Nesvorn\'y et al. 2010). Many CC binaries have widely separated components which can become 
unbound as a result of small collisions (Petit \& Mousis 2004). In contrast, the equal-size 
binaries are nearly absent in the hot population. Instead, in the hot population, it is more 
common to have a small satellite orbiting a much larger primary. These 
moons are thought to have accreted around primaries from impact-generated disks 
(Canup 2005, Leinhardt et al. 2010).

The most straightforward interpretation of these differences is that collisions played an important role
in shaping the hot population, whereas the collisional evolution of CCs was relatively modest
(Nesvorn\'y et al. 2011, Parker \& Kavelaars 2012). The collisional activity in the present 
Kuiper belt is low and not very different between the hot and cold populations. This means 
that the hot population must have collisionally evolved {\it before} it was implanted into 
the Kuiper belt, probably during a stage when it was embedded in the massive planetesimal disk. 
CCs did not follow the same evolution path most likely because the outer extension of the planetesimal disk 
at $>$40 au had a relatively low mass. In addition, it has been pointed out (Parker \& Kavelaars 2010)
that the wide binaries in the CC population would not dynamically survive during the implantation 
process, thus strengthening the idea that they formed beyond $\simeq$40~au.

The goal of this work is to establish the survival rate of binaries that formed in the massive 
planetesimal disk below 30 au.\footnote{A handful of equal-size binaries have been detected on highly 
inclined heliocentric orbits in the dynamically hot population (e.g., 2002 VF130, 2004 PB108; Section 3).
Their existence indicates that the formation of equal-size binaries was widespread and also occurred 
the massive disk below 30 au, which is thought to be the main source of the hot population. Indeed, 
the suggested binary formation mechanisms show only a weak dependence on the radial distance.}
First, before their implantation into the Kuiper belt, binaries 
can become unbound by kicks induced by small impacts. Second, during their implantation in the Kuiper belt, 
binaries have close encounters with migrating Neptune. Wide binaries can become unbound if Neptune's 
tidal potential during an encounter exceeds the binary binding energy (Parker \& Kavelaars 2010). 
More tightly bound binaries may have survived because their binding energy is greater. In addition, impacts  
and dynamical perturbations can change binary orbits, leading in some cases to a low-speed collision between 
binary components (Nesvorn\'y et al. 2018). This could potentially represent an interesting channel 
for the formation of {\it contact} binaries (e.g., Sheppard \& Jewitt 2004, Thirouin \& Sheppard 2018).

Binary survival in the massive planetesimal disk depends on the disk's lifetime, $t_{\rm disk}$. Clearly, 
if the disk was short-lived ($t_{\rm disk}<100$ Myr), fewer collisions would have occurred, and binaries 
would have had a better chance of survival than in the case of a long-lived disk ($t_{\rm disk}>100$~Myr). The disk 
lifetime, in turn, is related to the timing of Neptune's migration. In addition, the dynamical survival 
of binaries during their implantation into the Kuiper belt depends on the number and nature of planetary 
encounters, and ultimately on the orbital behavior of planets. We thus see that the binary occurrence in 
the Kuiper belt can be used, at least in principle, to constrain Neptune's migration and the implantation 
process itself. 

In summary, the goal of this paper is to test a specific hypothesis: that binaries in transneptunian 
populations other than low inclination CCs originally formed at lower heliocentric distance and were 
transported to their current orbits by way of interactions with giant planets. This hypothesis is 
currently the most frequently asserted and is consistent with several observed features of the population, 
but many uncertainties remain, for example, related to the radial profile of the planetesimal disk.
Here we adopt an assumption that the source of the entire hot population was a massive planetesimal disk 
with an outer edge at $\sim$30 au. This remains a hypothesis. We find that the existence of very wide
binaries in the hot population may not be fully consistent with this hypothesis, possibly indicating that 
at least part of the hot population started beyond the reach of Neptune at $>$30 au.    

\section{Previous Work on Binary Survival}

The problem of KBO binary survival was considered in several publications. Petit \& Mousis (2004) pointed 
out that small impacts can dissolve a binary if the velocity change produced by impacts is comparable to 
binary's orbital speed (typically meters per second for known KBO binaries). The work determined the 
expected lifetime of several binaries in the current Kuiper belt environment (and discussed the effects of 
primordial grinding as well). Using a reference size distribution of impactors with $N(R) \propto R^{-4.5}$ 
for radius $R>5$ km and $N(R) \propto R^{-3}$ for $R<5$ km, they found that wide binaries such as 1998 
WW31 and 2001 QW322 have expected lifetimes of only $\sim$1-2 Gyr. This would suggest that these binaries 
have been much more common in the early Solar System (for a few to survive to the present day). 

The effect of impacts, however, depends on the assumed size distribution of impactors and much longer 
lifetimes are inferred if $N(R) \propto R^{-3}$ for $5<R<50$ km, as suggested by modern observational surveys, 
Charon craters and other constraints (e.g., Bernstein et al. 2004, Morbidelli et al. 2009, Parker \& Kavelaars 
2012, Fraser et al. 2014, Nesvorn\'y et al. 2018, Singer et al. 2019). Additional mechanisms studied in 
Petit \& Mousis (2004), such as collisional disruption and gravitational scattering by large KBOs, were found 
to be less of an issue for binary survival. Shannon \& Dawson (2018) modeled the effect of gravitational 
scattering in detail and concluded that the existence of wide CC binaries is consistent with 
$\sim$1000-4000 Pluto-mass objects in the original disk (Nesvorn\'y \& Vokrouhlick\'y 2016).   

Nesvorn\'y et al. (2011) extended the work of Petit \& Mousis (2004) by modeling impacts in the early 
Solar System. They showed that the existence of CC binaries can be used to set limits on the extent
of collisional grinding in the primordial Kuiper belt and suggested that the observed rollover of CCs with 
$R>50$ km (e.g., Fraser et al. 2014) was probably not produced by disruptive collisions. Instead, it may be 
a fossil remnant of the KBO formation process. In contrast, the size distribution break at $R \sim 50$ km 
in the hot population was most likely produced by collisional grinding during the early stages (e.g., 
Nesvorn\'y et al. 2018). Here we consider the effects of primordial grinding on binaries found the hot 
population and show that some of them (e.g., 2002 VF130) are expected to have a very low survival 
probability. This has interesting implications for the initial binary fraction and/or for the original 
source reservoir of binaries now found in the hot population. 

The {\it dynamical} survival of binaries found in the cold population was studied in Parker \& Kavelaars 
(2010) and Fraser et al. (2017). Parker \& Kavelaars (2010) showed that the large binary fraction in 
the cold population is inconsistent with them being implanted to $\simeq$42-47 au from a massive disk below $\sim$35 au 
(Levison et al. 2008). This is because bodies starting below $\sim$35 au often have scattering encounters 
with Neptune before reaching $\simeq$42-47 au, and most wide binaries are dissolved in the process. The binary contraint 
therefore gives a strong support to the idea that the CC population formed beyond $\sim$35 au (see Section 5.1 
for additional discussion of Parker \&  Kavelaars 2010). Fraser et al. (2017) complemented these results by 
demostrating that some wide CC binaries (presumably the ones with less red colors) could have been pushed 
from $\sim$38-42 au to $>$42 au by the 2:1 resonance with migrating Neptune. 

Here we study the effects of planetary encounters on binaries starting in the massive disk below $\sim$30 
au and reaching orbits in the hot population (i.e., HCs, resonant and scattered objects). This scientific 
problem was not considered before (see Noll et al. 2006 and Nesvorn\'y et al. 2018 for related studies of 
Centaurs and Jupiter Trojans) 
at least partly because we did not have a reliable model for the implantation of objects in the Kuiper belt. 
Here we use the implantation model of Nesvorn\'y \& Vokrouhlick\'y (2016), which has a long heritage in 
the previous works on the subject (e.g., Hahn \& Malhotra 2005, Levison et al. 2008). The model of
Nesvorn\'y \& Vokrouhlick\'y (2016) with slow and grainy migration of Neptune was shown to match the 
observed orbital structure of the Kuiper belt (inclination distribution, ratio of resonant and non-resonant 
objects, etc). Some of the model predictions (Kaib \& Sheppard 2016, Nesvorn\'y et al. 2016) have already 
been confirmed from new observations (e.g., Lawler et al. 2019). 

\section{Known Kuiper Belt Binaries}

\begin{figure}
\epsscale{0.8}
\plotone{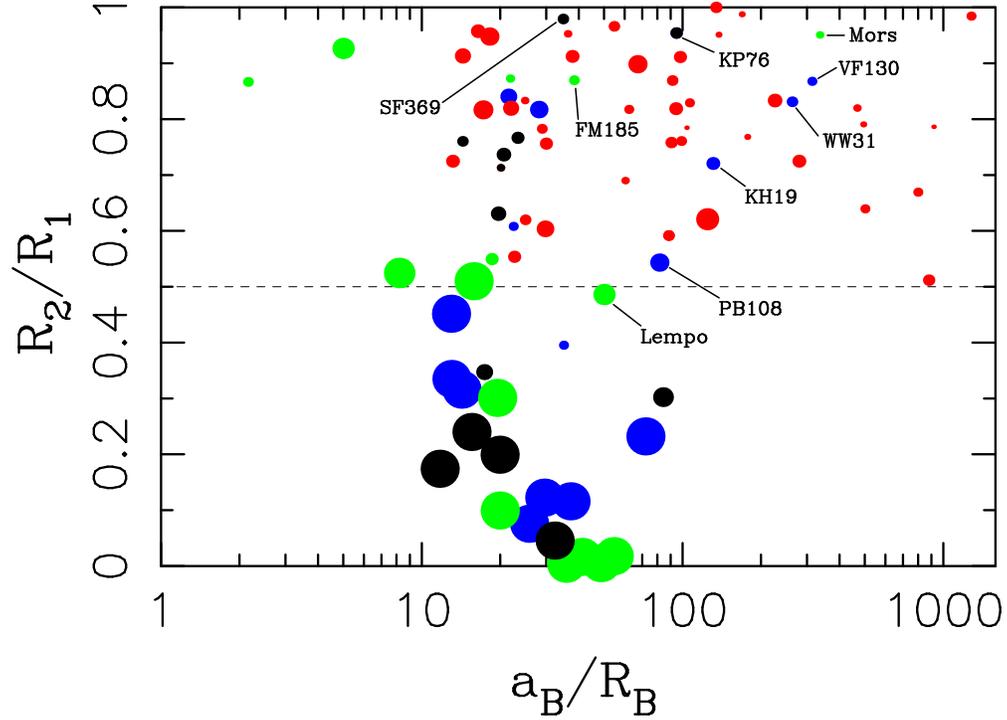}
\caption{The properties of KBO binaries. The color code indicates the relationship of binaries to different 
dynamical classes (red for CCs, green for Plutinos, blue for HCs, black for everything else; see Fig. 1). The symbol 
size correlates with the primary diameter. The unequal-size binaries with a large primary and a small moon 
($R_2/R_1<0.5$) are detected around large primaries in the dynamically hot populations. Most known 
equal-size binaries ($R_2/R_1>0.5$) are in the cold population.}
\label{realbin}
\end{figure}

A catalog of physical and orbital properties of binary bodies is maintained W. R. Johnston (Johnston 2018) 
on the NASA Planetary Data System (PDS) node.\footnote{{\tt https://sbn.psi.edu/pds/resource/binmp.html}} 
We analyzed the PDS catalog in September 2018. Figure \ref{realbin} shows the basic properties of 
KBO binaries/satellites. Several notable features are apparent in the plot. First, 
the unequal-size binaries with a large primary and a small moon ($R_2/R_1<0.5$, where $R_1$ and $R_2$
denote the primary and secondary radii) are mainly detected around large primaries in the hot population. 
They are absent in the cold population either because they did not form or because bodies in the CC population 
are generally smaller and the moons with $R_2<0.5\,R_1$ around small primaries are difficult to detect.
Second, most known equal-size binaries with $R_2/R_1>0.5$ appear in the cold population (40 out of 65 known;
shown in red in Fig. \ref{realbin}). 

Two (1998 WW31 and (119067) 2001 KP76) of only seven known HC binaries (see Section 4.1 for our definition 
of dynamical categories) with $R_2/R_1>0.5$ have the heliocentric 
inclinations slightly above our cutoff limit of 5$^\circ$ and may be interlopers from the CC population (Fraser 
et al. 2017). This applies to (341520) Mors-Somnus and (82157) 2001 FM185, both in the Plutino population, as 
well.\footnote{(82157) 2001 FM185 with $a=38.7$ au was classified as CC by the Deep Ecliptic Survey. Here we
prefer to relate it to the 3:2 resonance ($a=39.4$ au) and include it in the present analysis.} 
 
Of the remaining five, 2002 VF130 has the largest binary separation ($a_{\rm B}/R_{\rm B}\simeq310$, 
where $a_{\rm B}$ is the binary semimajor axis and $R^3_{\rm B}=R_1^3+R_2^3$), and 2004 PB108 and 2004 KH19 have $a_{\rm B}/R_{\rm B} 
\simeq81$ and 130, respectively (Table 1). These wide binaries in the HC population represent the most interesting constraint. 
All other nearly equal-size binaries in the HC population have $a_{\rm B}/R_{\rm B}\lesssim30$ and better 
odds of survival (Section 5). In addition, (47171) Lempo (provisory designation 1999 TC36) is a triple system in the 
Plutino population with $R_3/R_1 \simeq 0.5$ and $a_{\rm B}/R_{\rm B} \simeq 50$, where $R_3$ is the radius of the 
tertiary component. The scattered disk binary 2006 SF369 has $R_2/R_1\simeq0.98$ and $a_{\rm B}/R_{\rm B}\simeq35$.

\section{Method}

\subsection{Dynamical Effect of Planetary Encounters}

We make use of the simulations published in Nesvorn\'y \& Vokrouhlick\'y (2016). See this work for 
the description of the integration method, planet migration, initial orbital 
distribution of disk planetesimals, and comparison of the results with the orbital structure of the 
Kuiper belt. A shared property of the selected runs is that Neptune migrates outward by scattering 
planetesimals (Table 2). Planetesimals were initially distributed in a disk extending from just beyond 
the initial orbit of Neptune at 22 au to 30~au. The outer extension of the disk beyond 30~au was 
ignored, because various constraints indicate that a large majority of planetesimals started at $<$30 
au (e.g., Gomes et al. 2004). The simulations were performed with a modified version of the symplectic 
$N$-body integrator known as {\it Swift} (Levison \& Duncan 1994). 

All encounters of planetesimals with planets were recorded during these simulations. This was done by 
monitoring the distance of each planetesimal from Jupiter, Saturn, Uranus and Neptune, and recording 
every instance when the distance dropped below 0.5 $R_{{\rm Hill},j}$, where $R_{{\rm Hill},j}$ are the Hill radii 
of planets ($j=5$ to 8 from Jupiter to Neptune). We verified that the results do not change when 
more distant encounters are accounted for. 

We selected disk planetesimals that ended up in different KBO populations at the end of  
simulations ($t=4.5$ Gyr). Specifically, we defined the following four categories: HCs (semimajor axes 
$40<a<47$ au, perihelion distances $q>36$ au, orbital inclinations $i>5^\circ$), Plutinos (stable 
librations in the 3:2 resonance), scattering ($50<a<200$ au, 
$>$1.5~au change in $a$ in the last Gyr; Gladman et al. 2008) and detached ($50<a<200$ au, $<$1.5 au 
change) objects. We used a longer time interval than Gladman et al. (2008) to distinguish between
the scattering and detached populations. Our definition of the detached population is therefore
more restrictive. The CC population ($42<a<47$ au, $q>36$ au, $i<5^\circ$) was not considered 
here.\footnote{If the CC population formed at $a\simeq 40$ au, the CC binaries have not experienced planetary 
encounters (e.g., Fraser et al. 2017).} Note that the distinction between CCs and HCs based on a single 
inclination cutoff is somewhat arbitrary. 

Each selected planetesimal was assumed to be a binary object. We considered a range of 
binary separations ($1<a_{\rm B}/R_{\rm B}<2000$), initially circular orbits (binary orbit eccentricity 
$e_{\rm B}=0$), and a random distribution of binary inclinations ($i_{\rm B}$). In some instances, 
several clones with different binary inclinations vere assigned to each selected planetesimal to increase the 
statistics. Each binary was evolved through each recorded planetary encounter. We used the Bulirsch-Stoer (B-S) 
$N$-body integrator that we adapted from Numerical Recipes (Press et al. 1992). The center 
of mass of each binary planetesimal was first integrated backward from the time of the closest approach to 
3~$R_{\rm Hill}$. It was then replaced by the actual binary and integrated forward through the encounter 
until the planetocentric distance of the binary exceeded 3~$R_{\rm Hill}$.\footnote{The code is available
upon request from authors.} The final binary orbit was used as an initial orbit for 
the next encounter and the algorithm was repeated over all encounters. 

The B-S code monitored collisions between binary components. If a collision occurred, the integration was stopped 
and the impact speed and angle were recorded. A fraction of binaries became unbound. For the surviving 
binaries, we recorded the final values of $a_{\rm B}$, $e_{\rm B}$ and $i_{\rm B}$, which were then used to 
evaluate the overall change of orbits. After all integrations finished, we combined the individual 
runs into a statistical ensemble of possibilities. The results convey the dynamical survival probability 
of binaries in each KBO category. 

\subsection{Collisional Survival}

The mutual orbit of a binary can be affected by small impacts into its components (Petit \& Mousis 2004). 
Here we investigate this process with the collision code that we previously developed (Morbidelli et al. 2009, 
Nesvorn\'y et al. 2011). The code, known as {\it Boulder}, employs a statistical method to track the 
collisional fragmentation of planetesimal populations. It was developed along the lines of other published 
codes (e.g., Weidenschilling et al. 1997, Kenyon \& Bromley 2001). A full description of the {\it Boulder} code, 
tests, and various applications can be found in Morbidelli et al. (2009), Levison et al. (2009) and 
Bottke et al. (2010). Here we briefly highlight the main points and differences with respect to 
these publications.

For each collision, the code computes the specific impact energy $Q$ and the critical impact energy 
$Q^*_{\rm D}$ for catastrophic disruption (see Benz \& Asphaug (1999) for definitions). Based on 
the value of $Q/Q^*_{\rm D}$ and available scaling laws, it then determines the masses of the largest 
remnant and largest fragment, and the power-law size distribution of smaller fragments (e.g., Durda et 
al. 2007). The $Q^*_{\rm D}$ function in {\it Boulder} was set to be intermediate between the impact 
simulations with strong (Benz \& Asphaug 1999) and weak ice (Leinhardt \& Stewart 2009). To achieve
this, we multiplied $Q^*_{\rm D}$ from Benz \& Asphaug (1999) by a factor $f_Q$, where $f_Q=1$, 0.3 and 
0.1 was used in different tests. 

The main input parameters are: the (i) initial size distribution of simulated populations, (ii) intrinsic 
collision probability $P_{\rm i}$, and (iii) mean impact speed $v_{\rm i}$. 
As for $P_{\rm i}$ and $v_{\rm i}$, we performed two different tests. The first test was intended to replicate the 
collisional grinding of the massive planetesimal disk. In this case, we assumed that migrating Neptune 
removed the disk at $t_{\rm disk}$ after the dispersal of the protosolar nebula ($t_0$), and let the disk 
collisionally evolve over $t_{\rm disk}$. The dynamical state of the disk was taken from 
Levison et al. (2011). For example, at 300 Myr after $t_0$, the disk at 20-30~au is characterized by $P_{\rm i} \simeq 8 
\times 10^{-21}$ km$^{-2}$ yr$^{-1}$ and $v_{\rm i} \simeq 0.4$ km s$^{-1}$ (Morbidelli \& Rickman 2015).

The second set of tests with {\it Boulder} was done under the assumption that the outer planetesimal 
disk was dispersed by Neptune immediately after $t_0$ (i.e., $t_{\rm disk}=0$). The disk was assumed 
to have started dynamically cold ($e \simeq 0$ and $i \simeq 0$).
It was gradually excited after $t_0$, on a timescale of $\sim$10-30 Myr (Table 2), by migrating Neptune. 
The \"Opik algorithm (Wetherill 1967, Greenberg 1982) and the simulations reported in Nesvorn\'y
\& Vokrouhlick\'y (2016) were used to compute $P_{\rm i}$ and $v_{\rm i}$ as a 
function of time. We monitored the collision probabilities and impact velocities of the selected 
planetesimals (i.e., the ones that ended up in one of the considered KBO categories at $t=4.5$ Gyr) 
with all other planetesimals. The $P_{\rm i}$ and $v_{\rm i}$ values were computed each $\delta t$ 
by averaging over the selected planetesimals, where $\delta t = 1$ Myr during the initial stages, 
when $P_{\rm i}$ and $v_{\rm i}$ change quickly, and $\delta t=10$-100 Myr later on. 

The initial size distribution of the massive disk can be informed from the planetesimal formation 
models (e.g., Simon et al. 2017), but we considered other possibilities as well (Section 5.3). The 
final size distribution was required to match the shape of the size distribution of Jupiter Trojans
(see, e.g., Morbidelli et al. 2009, Nesvorn\'y et al. 2013, Fraser et al. 2014, Singer et al. 2019 
for a justification of this assumption), which is well characterized from observations down to at least 
3 km diameter, $D$ (Wong \& Brown 2015, Yoshida \&  Terai 2017). For $5\lesssim D \lesssim 100$ km, 
the cumulative size distribution $N(>\!\!D)$ is a power law $N(>\!\!D) \propto D^{-\gamma}$ with $\gamma 
\simeq 2.1$. Above $D \simeq 100$ km, the Jupiter Trojan size distribution bends to a much steeper slope 
($\gamma \sim 6$). 

To construct the size distribution of planetesimals in the massive disk, the Jupiter Trojan size 
distribution was divided by $P_{\rm JT}$, where $P_{\rm JT}=5\times10^{-7}$ is the Jupiter Trojan
capture probability determined in Nesvorn\'y et al. (2013) (this is a probability that an outer disk 
planetesimal ends up on a stable Jupiter Trojan orbit). This gives $\simeq 6 \times 10^9$ 
disk planetesimals with $D>10$ km or $\simeq 5 \times 10^7$ disk planetesimals with $D>100$ km. 
The total mass of the reconstructed population is $\simeq$20 $M_\oplus$, where $M_\oplus$ is the Earth 
mass, in agreement with the results of Nesvorn\'y \& Morbidelli (2012). 

\subsection{Binary Module}

The binary module in {\it Boulder} (Nesvorn\'y et al. 2011) accounts for small, non-disruptive impacts 
on binary components, and computes the binary orbit change depending on the linear momentum of impactors. 
For each impact, the change of orbital speed, $\mathbf{v}_{\rm B}=\mathbf{v}_2-\mathbf{v}_1$, where 
$\mathbf{v}_1$ and $\mathbf{v}_2$ are the velocity vectors of components, is computed from the conservation 
of the linear momentum. This gives 
\begin{equation}
\delta \mathbf{v}_{\rm B} = {m_{\rm i} \over m_2+m_{\rm i}}\left( {1 \over 2} \mathbf{v}_{\rm i} - {m_1 \over m_{\rm B}} \mathbf{v}_{\rm B}\right) 
\end{equation} 
for an impact on the secondary, and
\begin{equation}
\delta \mathbf{v}_{\rm B} = - {m_{\rm i} \over m_1+m_{\rm i}}\left( {1 \over 2} \mathbf{v}_{\rm i} + {m_2 \over m_{\rm B}} \mathbf{v}_{\rm B}\right) 
\end{equation} 
for an impact on the primary, where $m_1$ and $m_2$ are the primary and secondary masses, $m_{\rm B}=m_1+m_2$, and
$m_{\rm i}$ and $\mathbf{v}_{\rm i}$ are the impactor's mass and velocity. 

The first term in Eqs. (1) and (2) corresponds to the transfer of the linear momentum. The factor 1/2 stands for the 
contribution of impactor's linear momentum to the translational motion (as averaged over all impact geometries).
The rest of linear momentum is consumed by the spin vector change of the impacted binary component. Note 
that this assumes that all collisions are completely inelastic. 

The impact velocity vectors were assumed to be randomly oriented in the reference frames of binaries. We also 
factored in that impacts can happen at any orbital phase and averaged the binary orbit changes over the orientation 
and phase. The changes of orbital elements, $\delta a_{\rm B}$ and $\delta e_{\rm B}$, were computed from 
\begin{equation}
 {\delta a_{\rm B} \over a_{\rm B}} = \pm {1 \over \sqrt{3}} \, \frac{m_{\rm i} v_{\rm i}}{m_{\rm B} v_{\rm B}} \;  
\label{a}
\end{equation}
and 
\begin{equation}
 \delta e_{\rm B}  = \pm {1 \over 2} \sqrt{\frac{5}{6}} \, \eta \, \frac{m_{\rm i} v_{\rm i}}{m_{\rm B} v_{\rm B}} \; ,
\label{e} 
\end{equation}
where $v_{\rm i}$ and $v_{\rm B}$ are the moduli of $\mathbf{v}_{\rm i}$ and $\mathbf{v}_{\rm B}$, and $\eta^2=1-e_{\rm B}^2$. 
The $\pm$ sign in front of the right-hand sides indicates that the individual changes can be positive or 
negative. Equations (3) and (4) were implemented in the {\it Boulder} code. A similar expression can be 
obtained for inclinations (Nesvorn\'y et al. 2011), but we do not discuss the inclination changes here.

\section{Results}

\subsection{Dynamical Survival}

We first evaluated the dynamical effect of planetary encounters. Using methods described in 
Section 4.1, we determined how the survival probability depends on the size and separation of binary 
components. We found that the binary survival depends on $a_{\rm B}/R_{\rm B}$, where $R_{\rm B}=(R_1^3+R_2^3)^{1/3}$, 
and not on $a_{\rm B}$, $R_1$ and $R_2$ individually. This is a consequence of the binary dissociation condition 
described in Agnor \& Hamilton (2006). A binary with the total mass $m_{\rm B}=m_1+m_2$ can become unbound 
when the planetocentric Hill radius of the binary, $r_{\rm Hill, B}=q(m_{\rm B}/3 m_{\rm pl})^{1/3}$, where $q$ is 
the distance of the closest approach and $m_{\rm pl}$ is the planet mass, becomes smaller than the binary 
separation; that is $r_{\rm Hill, B} < a_{\rm B}$. This condition yields
\begin{equation}
{a_{\rm B} \over R_{\rm B}} > {1 \over 3^{1/3}} \left({\rho \over \rho_{\rm pl}}\right)^{\!\!\!1/3} 
\!\! \left( {q \over R_{\rm pl}} \right ) \ ,
\end{equation} 
where $R_{\rm pl}$ and  $\rho_{\rm pl}$ are the planet radius and density. Here we assumed that the primary 
and secondary components of binaries have the same density, $\rho$. For exactly equal-size binaries with
$R_1=R_2$, $R_{\rm B}=(R_1+R_2)/2^{2/3}$.  

The closest encounters with Neptune typically have $q/R_{\rm pl} \sim 40$ in our simulations, suggesting that 
binaries with $a_{\rm B}/R_{\rm B} > 30$ should often become dissociated. This closely corresponds to Fig. 
\ref{scaled}, where the survival probability for $a_{\rm B}/R_{\rm B} > 30$ is $<$50\%. 

The survival probability is a strong function of binary separation (Figure \ref{scaled}). The binaries
with small separations ($a_{\rm B}/R_{\rm B}<30$) are tightly bound together and have high survival probabilities 
(50-90\%). They are affected only during extremely close encounters to planets, which do not happen too often. 
The binaries with larger separations ($a_{\rm B}/R_{\rm B}>30$) are more likely to become dissolved. This is 
expected because the wide binaries suffer larger orbital changes during planetary encounters. Also, in about 
10\% of cases, binaries end up their existence during a collision between the binary components (the 
$N$-body code stops when a collision is identified), which may have interesting implications for the origin 
of contact binaries in the Kuiper belt (Section 6.1). 

The results shown in Figure \ref{scaled} were computed assuming the bulk density $\rho=1$ g cm$^{-3}$.
According to Eq. (5), the critical semimajor axis scales with $\rho^{1/3}$. Therefore, the surviving 
fraction curve shown in Figure \ref{scaled} would shift left by a multiplication factor of 0.79 for 
$\rho=0.5$ g cm$^{-3}$ and right by a multiplication factor of 1.26 for $\rho=2.0$ g cm$^{-3}$.
We confirmed this by simulating cases with different densities. 

The dynamical survival probabilities shown in Figure \ref{scaled} are lower than those reported 
in Parker \& Kavelaars (2010). They found a $\simeq$60\% survival probability for $a_{\rm B}/R_{\rm B}=200$, 
whereas we only find $\simeq$2.5\% probability. This is, in part, related to a much richer 
history of planetary encounters in our model with the slow migration of Neptune. Parker \& Kavelaars 
(2010), instead, assumed a strong instability case from Levison et al. (2008). Additional differences 
arise due to different selection criteria. In Figure \ref{scaled}, we selected all bodies that ended in 
the HC population at $t=4.5$ Gyr. Parker \& Kavelaars (2010) monitored each particle’s semimajor axis 
and eccentricity for 1 Myr after the instability and identified all candidates that passed through 
``CC-like'' orbits (which no longer have close encounters with Neptune). As explained in Levison et al. 
(2008), these candidates typically start with low orbital inclinations at $\simeq$30-34 au and their 
inclinations remain low because their experience fewer-than-average scattering encounters with Neptune. 
This can readily explain the difference between our Figure \ref{scaled} and Parker \& Kavelaars (2010), 
because fewer and/or more distant encounters imply better chances of binary survival. The minimum encounter 
distance to Neptune reported in their Figure 1 is $q\simeq0.1$ au, whereas we often have 
$q \simeq 0.01$~au.

For Plutinos, HCs and detached objects, all planetary encounters happen during the initial stages before 
objects are implanted onto stable orbits in the Kuiper belt. Only the scattered disk objects remain 
coupled to Neptune. For them, $>$90\% of the recorded planetary encounters happen within the first 
$\simeq$200 Myr of the simulation. This implies that binaries are typically removed early, or never, 
and the binary fraction in the scattering population does not change much in the last 4 Gyr (see 
Section 5.5 for a discussion of the Kozai resonance for binaries with $i_{\rm B} \sim 90^\circ$; Porter \& 
Grundy 2012).

Finally, we compare the results for different KBO populations (Figure \ref{pops}) and for different 
migration histories of Neptune (Figure \ref{cases}). Figure \ref{pops} shows that no significant 
differences are expected between different KBO populations. This is a consequence of the statistically 
similar histories of planetary encounters for bodies implanted into different populations. There are 
also no important differences between the two cases considered here with different timescales 
of Neptune's migration (Figure \ref{cases}). In case1 (Table 2), there are on average 12 (0.35) 
encounters within 0.1 au (0.01 au) to Neptune for each captured HC object. In case2, there are on 
average 10 (0.33) such encounters. The two cases are therefore expected to produce similar survival 
probabilities.

\subsection{Orbits of Surviving Binaries}

Planetary encounters act to change binary orbits. Figure \ref{sema} illustrates the relationship 
between the initial and final semimajor axes of the surviving binaries. The orbital changes are minimal 
for $a_{\rm B}/R_{\rm B}<30$, but increase with binary separation. For example, 90\% of surviving orbits that 
end up with $a_{\rm B}/R_{\rm B}=40$ start with $30<a_{\rm B}/R_{\rm B}<60$. For $a_{\rm B}/R_{\rm B}>100$, 
the orbital changes are major and the binary semimajor axis can change by more than a factor of 2. 

Figure \ref{ae}a shows the distribution of final separations for binaries started with $a_{\rm B}/R_{\rm B}=50$. 
Whereas most orbits remain with the semimajor axes near the original one, the distribution also shows 
wide wings toward lower and higher values. For example, $\simeq$10\% of surviving orbits end up with 
$a_{\rm B}/R_{\rm B}>70$. This shows that at least some of the wide KBO binaries in the hot populations 
could have started with tighter orbits. 

The binary eccentricity changes can be substantial as well (Figure \ref{ae}b). For example, the 
mean eccentricity of the surviving orbits with initial $a_{\rm B}/R_{\rm B}=50$ is $e_{\rm B}\simeq0.3$
(this assumes that $e_{\rm B}=0$ initially). Thus, even if the equal-size KBO binaries in the 
dynamically cold and hot population presumably formed by the same mechanism, and initially
had the same distribution of binary orbits, they are not expected to be the same today. 
 
\subsection{Collisional Survival of Binaries in the Massive Disk}

The collisional evolution of the massive outer disk was studied in Nesvorn\'y et al. (2018). 
They showed that the Patroclus-Menoetius (P-M) binary in the Jupiter Trojan population poses an 
important constraint on the massive disk lifetime ($t_{\rm disk}$). This is because the longer the
P-M binary stays in the disk, the greater is the likelihood that its components will be stripped 
from each other (by the impact-related process described in Section 4.3; Petit \& Mousis 2004). 
They found that the 
massive disk must have been dispersed by the migrating planets within $\sim$100~Myr after the 
removal of the protosolar nebula (i.e., $t_{\rm disk} \lesssim100$~Myr). Here we first briefly 
recall their results related to the collisional grinding of the massive disk.

The collisional grinding of the outer planetesimal disk proceeds fast. For $t_{\rm disk}>100$~Myr, 
the number of $D > 10$ km bodies is reduced at least ten times and the total mass drops to 
$<$10 $M_\oplus$. These results are in conflict with the current size distribution of Jupiter 
Trojans (Wong \& Brown 2016), the planetesimal disk mass inferred from the Jupiter Trojan capture 
(Morbidelli et al. 2005, Nesvorn\'y et al. 2013), and other constraints (e.g., Nesvorn\'y \& 
Morbidelli 2012). The problem could 
potentially be resolved if a larger initial mass was adopted. Nesvorn\'y et al. (2018) tested
several possibilities along these lines. For example, they scaled up the reference size distribution 
by an additional factor to increase the initial mass to $>$20 $M_\oplus$. These tests failed because 
more massive disks grind faster and end up with $<$10 $M_\oplus$ for $t_{\rm disk}>100$ Myr. In other 
tests, they used a steeper slope for $D<100$ km in an attempt to obtain $\gamma \simeq 2$ as a 
result of collisional grinding. These tests failed as well for reasons similar to those described 
above. 

Given these unresolved issues, we decided to adopt the following scheme for our nominal simulations 
of impacts on KBO binaries. We used the reference size distribution (20 $M_\oplus$ initially) and 
switched off the fragmentation of planetesimals ($f_Q \gg 1$). In this case, the size distribution 
stayed approximately the same over the whole length of the simulation. This is arguably a very 
conservative assumption. Other schemes would require that the initial population was larger and 
decayed over time, implying more impacts overall. 

Figure \ref{col} shows the survival of equal-size binaries in our nominal simulations. The survival 
probability is a strong function of the physical size of binary components, their separation, and
$t_{\rm disk}$. Panel (b) is similar to Fig. 2 in Nesvorn\'y et al. (2018), where the results were 
obtained for the P-M binary ($R_1+R_2=109$ km and $\rho=0.88$ g cm$^{-3}$; here we use $R_1+R_2=100$ 
km and $\rho=1$ g cm$^{-3}$). It shows that the equal size binaries with $R_1+R_2=100$ km are not 
expected  to survive, on average, unless $t_{\rm disk}<100$ Myr. Binaries with smaller (larger) components 
have lower (higher) survival chances, as already noted in Nesvorn\'y et al. (2011) and Parker et 
al. (2012). The implications of these results for the binary occurrence in the Kuiper belt are 
discussed in Section 6. 

Motivated by the results of Nesvorn\'y et al. (2018), we now consider $t_{\rm disk}<100$ Myr 
(Figures \ref{si} and \ref{si2}). The rounded shape of the initial size distribution was informed 
from the hydrodynamical simulations of the streaming instability (Simon et al. 2017). The 
characteristic size of planetesimals formed by the streaming instability was set to $D=100$~km. 
In Fig. \ref{si}, we used the initial disk mass $M_{\rm disk}=30$ $M_\oplus$ and $f_{\rm Q}=0.1$,
and let the disk grind for $t_{\rm disk}=10$ Myr. In Fig. \ref{si2}, we adopted $M_{\rm disk}=40$ 
$M_\oplus$, $f_{\rm Q}=0.3$ and $t_{\rm disk}=50$ Myr. In both these cases, the final size distribution 
was required to fit the size distribution reconstructed from Jupiter Trojans (Section 4.2) and, 
indeed, Figures \ref{si} and \ref{si2} show that both simulations satisfied this constraint 
quite well.   

The binary survival probability is significantly lower in the case of larger initial disk 
mass and longer disk lifetime. This is most obvious for binaries with smaller components,
$R_1+R_2=30$ km, for which the survival for $M_{\rm disk}=40$ $M_\oplus$ and $t_{\rm disk}=50$ Myr 
is roughly 10 times lower than for $M_{\rm disk}=30$ $M_\oplus$ and $t_{\rm disk}=10$ Myr (compare 
Figs. \ref{si}b and \ref{si2}b). The lower survival probability is expected because more massive 
disks with longer lifetimes provide more impactors overall. More massive binaries have better chances 
of survival. For example, for $R_1+R_2=300$ km and $a_{\rm B}=10,000$ km, the survival probability 
is $\simeq$45\% for $M_{\rm disk}=40$~$M_\oplus$ and $t_{\rm disk}=50$ Myr, and $\simeq$80\% for 
$M_{\rm disk}=30$ $M_\oplus$ and $t_{\rm disk}=10$ Myr. 

The two cases described here (Figs. 10 and 11) are not unique. In fact, there is a continuous range of parameters 
(initial $M_{\rm disk}$, $t_{\rm disk}$, $f_Q$, etc.) that satisfy the existing constraints (mainly
the size distribution inferred from Jupiter Trojans). It is therefore unclear at this point
whether the binary survival was higher, such as in Fig. \ref{si}, or lower, such as in Fig. 
\ref{si2}, or whether some other case not considered here could end up giving a different 
result. In any case, our results show how the binary occurrence in the hot population of the 
Kuiper belt is linked to the mass and lifetime of the massive planetesimal disk. Additional work
will be needed to resolve this relationship in more detail. 

\subsection{Collisional Survival During Subsequent Epochs}

The massive planetesimal disk below 30 au is dispersed when Neptune migrates into it. A small fraction 
of disk planetesimals is subsequently implanted into different populations in the Kuiper belt. The exact 
timing of these events is uncertain, but the P-M binary constraint implies that $t_{\rm disk}<100$ Myr 
(Nesvorn\'y et al. 2018). Examples with $t_{\rm disk}=10$ and 50 Myr were discussed in the previous section. 
Here we adopt $t_{\rm disk}=0$ and consider a case when Neptune migrates into the planetesimal disk 
immediately after $t_0$. The impact probability $P_{\rm i}$ and $v_{\rm i}$ were evaluated as a function of 
time as described in Section 4.2 (Figure \ref{varcol}). The changing conditions were implemented in 
the {\it Boulder} code, which was then used to determine the collisional survival of binaries over 
the past 4.5 Gyr. 

We found that, to fit the present size distribution of Jupiter Trojans, the shape of the size distribution 
at $t_0+t_{\rm disk}$ must have been similar to the present one for $D>10$ km (Figure \ref{var1}a). This
is because the impact probability drops relatively fast such that not much grinding happens for 
$D>10$ km over 4.5 Gyr (with $t_{\rm disk}=0$). The surviving fraction of equal-size binaries is shown in Figure \ref{var1}b. 
The survival probability is sensitive to the size of binary components and their separation. 
For $R_1+R_2=100$-300 km, which is the characteristic size of known equal-size binaries, the 
survival probability is 74-96\% for $a_{\rm B}\simeq$1,000-10,000 km. Most of these binaries are
therefore expected to survive. A sharp drop-off of the probability at $a_{\rm B}/(R_1+R_2) \sim 10^3$
occurs because separations approach 0.5 $R_{\rm B,Hill}$, where $R_{\rm B,Hill}$ is the binary heliocentric 
Hill radius (e.g., $R_{\rm B,Hill}\simeq250,000$ km for an equal size-binary with $R_1+R_2=100$ km,
$\rho=1$ g cm$^{-3}$ and $a=30$ au). These very wide binaries are dynamically unstable.  

The results shown in Figures \ref{varcol} and \ref{var1} were obtained for the case2 migration parameters 
(Table 2) and HCs, but the results for case1 and other KBO populations are similar. The survival probability of 
binaries is sensitive to our assumption about the number of disk planetesimals at $t_0+t_{\rm disk}$,
which is linked, through the implantation probability, to the number of KBOs. We showed previously
(e.g., Nesvorn\'y \& Vokrouhlick\'y 2016, Nesvorn\'y et al. 2016) that our baseline dynamical model 
for the Kuiper belt implantation produces populations that are similar to those inferred from 
observations of HCs, Plutinos, scattering and detached population. Still, there is some uncertainty 
in this comparison both on the side of observations (albedo assumptions, survey biases, etc.) and
model (dependence on the timescale of Neptune's migration, number of Pluto-size bodies in the disk, 
etc.). The initial calibration of the disk population is also uncertain, to within a factor of 
$\sim$2, because the capture probability of Jupiter Trojans depends on various details of the 
model setup.

To tentatively account for these uncertainties, we considered cases with $M_{\rm disk}=10$~M$_\oplus$
and 30 M$_\oplus$ (recall that $M_{\rm disk}=20$ M$_\oplus$ at $t_0+t_{\rm disk}$ in our nominal case).
This would correspond to $P_{\rm JT}=7.5\times10^{-8}$ and $2.5\times 10^{-8}$, respectively 
(see Section 4.2), which is somewhat outside the range of values favored in Nesvorn\'y et al. (2013),
but still potentially plausible. By testing these cases, we found that the survival probabilities 
of equal size binaries are very similar (within $\sim$20\%) to those shown in Figure \ref{var1}b. 
The results shown in Fig. \ref{var1}b are therefore relatively robust. 

In summary, the overall survival 
of equal-size binaries mainly depends on the early phase, when the binaries are immersed in the massive 
planetesimal disk and have to withstand an intense initial bombardment. The magnitude of the initial
bombardment depends on $M_{\rm disk}$ and $t_{\rm disk}$. The occurrence of binaries in the Kuiper belt 
could, at least in principle (i.e., after factoring in the effect of planetary encounters; Section 5.1), 
be therefore used to constrain $M_{\rm disk}$ and $t_{\rm disk}$ (see Section 6 for discussion).
    
\subsection{Kozai Dynamics and Tides}

Additional effects that can alter binary orbits, and may therefore influence binary survival, include Kozai 
dynamics and tides (e.g., Porter \& Grundy 2012). The Kozai dynamics of a binary orbit arises due to the 
gravitational potential of the Sun. For the Kozai cycles to be effective, the binary components must be 
roughly spherical and/or the binary separation must be large. If not, the gravitational potential from 
the $J_2$ term of binary components prevails over the solar gravity, resulting in a simple precession 
of the binary orbit about the heliocentric orbit pole. 

Porter \& Grundy (2012) studied these effects, including the tidal dissipation (Goldreich \& Sari 2009), 
and concluded that the combined effect of Kozai cycles and tides can remove binaries with inclinations 
$i_{\rm B} \sim 90^\circ$ and $a_{\rm B}>a_{\rm crit}$, where $a_{\rm crit}$ depends on $J_2$ and the strength of 
tidal dissipation. Given that several known Kuiper belt binaries with $a_{\rm B}<0.05$~$R_{\rm Hill}$ have 
$i_{\rm B} \sim 90^\circ$, this probably implies $a_{\rm crit}>0.05$ $R_{\rm Hill}$. This reasoning, if applied 
to the Kuiper belt binaries with $a_{\rm B}<0.05$ $R_{\rm Hill}$, suggests that these additional effects 
do not play a major role in their survival. The Kozai cycles and tides probably acted to eliminate 
binaries with $a_{\rm B}>0.05$ $R_{\rm Hill}$ and  $i_{\rm B} \sim 90^\circ$ (Grundy et al. 2018).

\subsection{Large Primaries with Small Satellites}

Our base hypothesis is that the (nearly) {\it equal}-size binaries with $R_2/R_1>0.5$ formed during the 
earliest stages (Goldreich et al. 2002, Nesvorn\'y et al. 2010). In contrast, it is not quite 
clear when the {\it unequal}-size binaries with $R_2/R_1<0.5$ formed. They are presumably a by-product of large scale 
collisions (Canup 2005, Leinhardt et al. 2010). On one hand, the collisional activity is the 
most intense during the lifetime of the massive disk, before the disk is dispersed by Neptune. 
On the other hand, there is at least one case where there are good reasons to believe that the 
satellite-forming collision happened relatively late, that is {\it after} the massive disk 
dispersal. This is the case of two Haumea moons, Hi'iaka and Namaka. Haumea has a collisional family 
(Brown et al. 2007), which must have formed after the implantation of Haumea onto its current orbit 
(the family would otherwise be dispersed during implantation). Thus, if Hi'iaka and Namaka formed 
as a result of the family-forming collision, their formation most likely post-dates the epoch of Neptune's migration.  

For unequal-size binaries that formed before the disk dispersal, the dynamical survival probability
can be inferred from Figure \ref{scaled}. For example, Charon, Styx and Hydra have $a_{\rm B}/R_{\rm B}=15.8$, 
34.4 and 52.3, respectively.\footnote{For Styx and Hydra, we define $R_{\rm B}^3=R_1^3+R_2^3$, where $R_1$
is the effective radius of Pluto and Charon ($\simeq$1238 km).} According to Fig. \ref{scaled}, we find 
that these moons are expected to survive in 67\%, 45\% and 31\% of cases. In Fig. \ref{scaled}, however, 
we assumed $\rho=1$ g cm$^{-3}$, which is not adequate for Pluto ($\rho=1.9$ g cm$^{-3}$) and Charon 
($\rho=1.7$ g cm$^{-3}$). We therefore opted for simulating the survival of Pluto's moons directly. 
The results are given in Table 3 (they are roughly consistent with shifting the lines in 
Fig. \ref{scaled} by a factor of $(1.8)^{1/3}$ to the right). Charon survives in 91\% of trials in the 
case2 simulation and in 94\% of trials in the case1 simulation. This is comforting (see also Pires et al. 2015).

For Hydra, which is the farthest of Pluto's known moons, the survival probabilities are 74\% in case2 
and 82\% in case1. Also, in 41\% and 53\% of trials, respectively, Hydra ends up with the final orbital 
eccentricity below 0.01 (this assumes that is started with $e=0$). For Styx, which is the closest of 
Pluto's small moons, the eccentricity stays below 0.01 in 56\% and 66\% of trials. For comparison, the 
present orbital eccentricities of Hydra and Styx are $\simeq$0.006 (Showalter \& Hamilton 2015). It is plausible 
that these eccentricities were generated by perturbations during encounters of the Pluto system to Neptune,
but the probability of that happening is small (Table 3). In addition, it is possible that Hydra started 
in the 6:1 resonance with Charon and was displaced from that resonance by planetary encounters. This requires 
a negative semimajor axis change $\Delta a \simeq -140$ km. For comparison, we find that $\Delta a < 
-100$ km in only $\sim$15\% of cases.

Eris and Dysnomia with $a_{\rm B}/R_{\rm B}\simeq31$ show larger survival probabilities 
than Styx (Eris's density is $\rho=2.5$ g cm$^{-3}$). Makemake's satellite, known as MK2, has 
a large orbital uncertainty with semimajor axis ranging between 21,000 and 300,000 km (Parker et al. 2016). 
This would give $a_{\rm B}/R_{\rm B}\simeq29$-410. The cases with $a_{\rm B}/R_{\rm B}>100$ can probably be 
ruled out because the survival odds are $\lesssim$10\% for these large separations. Quaoar-Weywot and 
Orcus-Vanth have $a_{\rm B}/R_{\rm B}\simeq26$ and 19, respectively, and are safe ($>$50\% survival probabilities 
according to Figure \ref{scaled}, but this is a generous lower limit given that, for example, Orcus density
is $\rho=1.5$ g cm$^{-3}$). In summary, we find that all known satellites of the largest KBOs are likely
to survive during the dynamical implantation of these bodies in the Kuiper belt.

We performed several tests to understand the effect of impacts on unequal size binaries. For example, 
we considered the small satellites of Pluto during and after the phase of the massive disk dispersal
(i.e., the case with $t_{\rm disk}=0$). These tests closely followed the setup described in Section 5.4 (see Fig. 
\ref{var1}a). All moons were started on circular orbits. We found that the survival probability 
of all small Pluto moons is very nearly 100\%. As for Hydra, the characteristic change of the semimajor 
axis and eccentricity due to impacts is only $\sim$100~km and 0.001, too low to explain Hydra's 
current orbital eccentricity ($\simeq$0.006). The results for Kerberos are similar, but Kerberos's 
eccentricity is smaller ($\simeq$0.003). For example, in about 30\% of cases, Kerberos ends up with 
$e>0.0015$ (i.e. at least a half of it present eccentricity). Larger effects are expected for $t_{\rm disk}>0$. 
This shows that at least some of Pluto's moons (not Charon) may own their slightly excited orbits 
to small impacts. 
    
\section{Discussion}

\subsection{Contact and Very Tight Binaries} 

The equal-size ($R_2/R_1>0.5$) binaries in the hot population can be divided into several categories. 
The contact and very tight binaries with $a_{\rm B}/R_{\rm B}<10$, such as 2001 QG298 and the inner pair 
of (47171) Lempo, are difficult to detect observationally. Here we showed that $\simeq$10\% of 
equal-size binaries are expected to collapse into contact binaries during the implantation of objects
into the Kuiper belt (Fig. \ref{scaled}). In addition, $\sim$10-30\% of equal-size binaries are 
expected to collapse into contact binaries during the collisional evolution of the massive disk
(Nesvorn\'y et al. 2018). Assuming a 100\% initial binary fraction, these results imply that the 
fraction of contact binaries in the hot 
population should be of the order of 10-30\%. For comparison, Sheppard \& Jewitt (2004) proposed
from the detection of 2001 QG298 (contact binary in the Plutino population) that at least 
$\sim$10\% of KBOs are contact binaries, and Thirouin \& Sheppard (2018) suggested that up to $\sim$40\% 
Plutinos can be contact binaries. The contact binary fraction among CCs remains to be determined, 
but if the main channel of contact binary formation is binary collapse from impact and planetary 
encounter perturbations, the contact binary fraction among $\sim$100-km-class CCs is expected to be 
low (most large CC binaries presumably survived).  Recent observations of (486958) 2014 MU69 by 
the New Horizons spacecraft show that contact binaries may be common among smaller, 
$\sim$10-km-class CCs. 

\subsection{Tight Binaries}

The tightly-bound binaries with $10<a_{\rm B}/R_{\rm B}<30$ have $>$50\% dynamical survival probability 
(Fig. \ref{scaled}) and do not provide a useful constraint on the implantation process itself. The 
nine known binaries with $10<a_{\rm B}/R_{\rm B}<30$ represent a fraction of $\sim$3\% of dynamically 
hot KBOs that were searched for binaries with the Hubble Space Telescope (HST; W. Grundy, personal 
communication; roughly 300 hot KBOs were imaged by the HST). Some tight binaries may 
remain unresolved. Thus, as a lower limit, we can estimate
that at least $\sim$6\% of the massive disk planetesimals were equal-size binaries with $10<a_{\rm B}/R_{\rm B}<30$. 
This is similar to the fraction of $a_{\rm B}/R_{\rm B}<10$ binaries inferred from the P-M binary constraint 
(Nesvorn\'y et al. 2018). Moreover, since the initial fraction of $10<a_{\rm B}/R_{\rm B}<30$ 
binaries cannot exceed 100\%, their survival probability during the collisional grinding of the 
massive planetesimal disk cannot be much lower than $\sim$0.06. This constraint implies $t_{\rm disk}<100$ 
Myr (Fig. \ref{col}). It means that the massive planetesimal disk must have been dispersed by 
migrating planets within $\lesssim$100 Myr after $t_0$ (Nesvorn\'y et al. 2018).

\subsection{Wide Binaries}

The wide binaries with $30<a_{\rm B}/R_{\rm B}<100$ have $\sim$10-50\% dynamical survival probability 
(Fig. \ref{scaled}). There are five known binaries in this category in the hot population (2001 KP76,
2004 PB108, 2006 SF369, 2001 FM185 and the outer component of (47171) Lempo; Table~1), representing a 
fraction of $\sim$2\% of all HST targets in the hot population. This implies the original 
fraction of at least $\sim$6\%, which is similar to the minimal binary fractions found for
$a_{\rm B}/R_{\rm B}<10$ and $10<a_{\rm B}/R_{\rm B}<30$. Together, we infer that at least $\sim$18\% of the massive 
disk planetesimals were equal-size binaries with $a_{\rm B}/R_{\rm B}<100$. Lower fractions 
would be inferred from the present analysis if 2001 KP76, 2001 FM185, and possibly (47171) Lempo,
turn out to be interlopers from the CC population (these objects have small heliocentric 
inclinations and may have evolved to their current orbits from $a>30$ au). 

\subsection{Very Wide Binaries}

Finally, there are four known equal-size binaries in the hot population with $a_{\rm B}/R_{\rm B}>100$
(1998 WW31, 2002 VF130 and 2004 KH19, all HCs, and Mors-Somnus in the Plutino population). These very 
wide binaries are puzzling because, according to Fig. \ref{scaled}, they have a very low chance of 
survival during their implantation into the Kuiper belt. 1998 WW31, however, has a heliocentric orbit 
with low inclination ($i\simeq7^\circ$) and we are therefore not confident whether it really formed 
in the massive disk below 30 au; it may be an interloper from the cold population. Similarly, (341520) 
Mors-Somnus with $i\simeq11^\circ$ may have been swept into the 3:2 resonance from its original 
formation location at $\sim$35-39 au (Nesvorn\'y 2015). If so, this binary could have avoided 
perturbations during planetary encounters. 

2004 KH19 and 2002 VF130, whose classification as hot KBOs is secure, only have 8\% and 0.8\% dynamical 
survival probabilities (Fig. \ref{scaled}). This could potentially imply a large fraction of the very wide 
binaries in the original disk. For example, from 2002 VF130, we could infer that the fraction of 
equal-size binaries with $a_{\rm B}/R_{\rm B}\sim300$ in the original disk was $\sim$40\%. 
And that's without factoring in the survival probability during the collisional grinding of the 
massive disk (Fig. \ref{col}b) in which 2002 VF130 presumably formed.

\subsection{Comparison of Cold and Hot Binaries}

Figure \ref{model} offers a different perspective on the same issue. There we compare the separations
of equal-size binaries in the cold and hot populations. The hot population shows a tail of contact and 
very tight binaries with $a_{\rm B}/R_{\rm B}<10$. These binaries are difficult to detect in the cold 
population, because of their greater heliocentric distance. If the binaries with $a_{\rm B}/R_{\rm B}<10$ 
are removed from the hot population, the separations of hot and cold binaries appear to follow the 
same trend, except that the cold population has nearly 20\% of equal-size binaries with $a_{\rm B}/R_{\rm B}>350$. 
These extremely wide binaries are missing in the hot population either because they did not form or because 
they were disrupted by impacts and planetary encounters. 

Now, the binaries with $100<a_{\rm B}/R_{\rm B}<350$ found in the hot population represent a problem, 
because they should have been removed by planetary encounters as well. To demonstrate this we use the 
equal-size binaries in the cold population as a template and factor in the dynamical survival probability 
from Fig. \ref{scaled}. The result is plotted as the ``model'' distribution in Fig. \ref{model}. Indeed, this 
simple test shows that the binaries with $100<a_{\rm B}/R_{\rm B}<350$ should have been nearly completely 
removed from the hot population, while in reality they represent $\sim$16\% of cases (4 out of 25). 

This may mean one of several things. It may indicate, for example, that at least some fraction of hot 
KBOs reached their current orbits without experiencing planetary encounters. For that to work, these 
KBOs would have to start beyond the reach of Neptune at $>$30 au, where they would only be affected 
by resonances with Neptune (Hahn \& Malhotra 2005), which are known to preserve binarity (Fraser 
et al. 2017). This scenario appears plausible for 1998 WW31 and (341520) Mors-Somnus, which have 
relatively low heliocentric orbit inclinations. It remains to be shown, however, whether this could 
explain 2002 VF130 ($i=19.5^\circ$) as well. Unfortunately, the binary semimajor axis of 2002 VF130,
estimated from the discovery image (Noll et al. 2009), remains uncertain and no additional astrometry 
is currently available. 

\section{Conclusions}

\begin{figure}
\epsscale{0.8}
\plotone{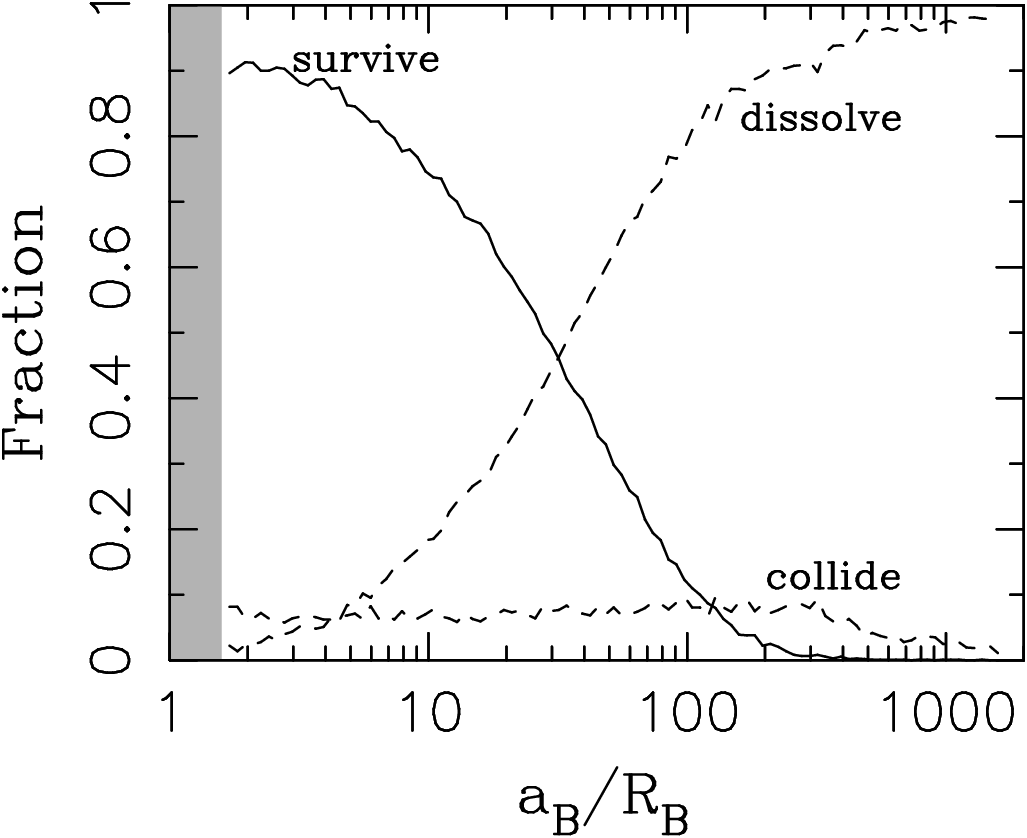}
\caption{The dynamical survival of equal-size binaries that end up in the HC population
(solid line). For example, 50\% of binaries survive for $a_{\rm B}/R_{\rm B}\simeq28$ and 10\% 
survive for $a_{\rm B}/R_{\rm B}\simeq110$. The dashed lines show the fractions of binaries 
whose components become unbound or end up colliding with each other. The gray area denotes 
separations for which the binary components are in contact ($a_{\rm B}/R_{\rm B}<2^{2/3}$).}
\label{scaled}
\end{figure}

\begin{figure}
\epsscale{0.8}
\plotone{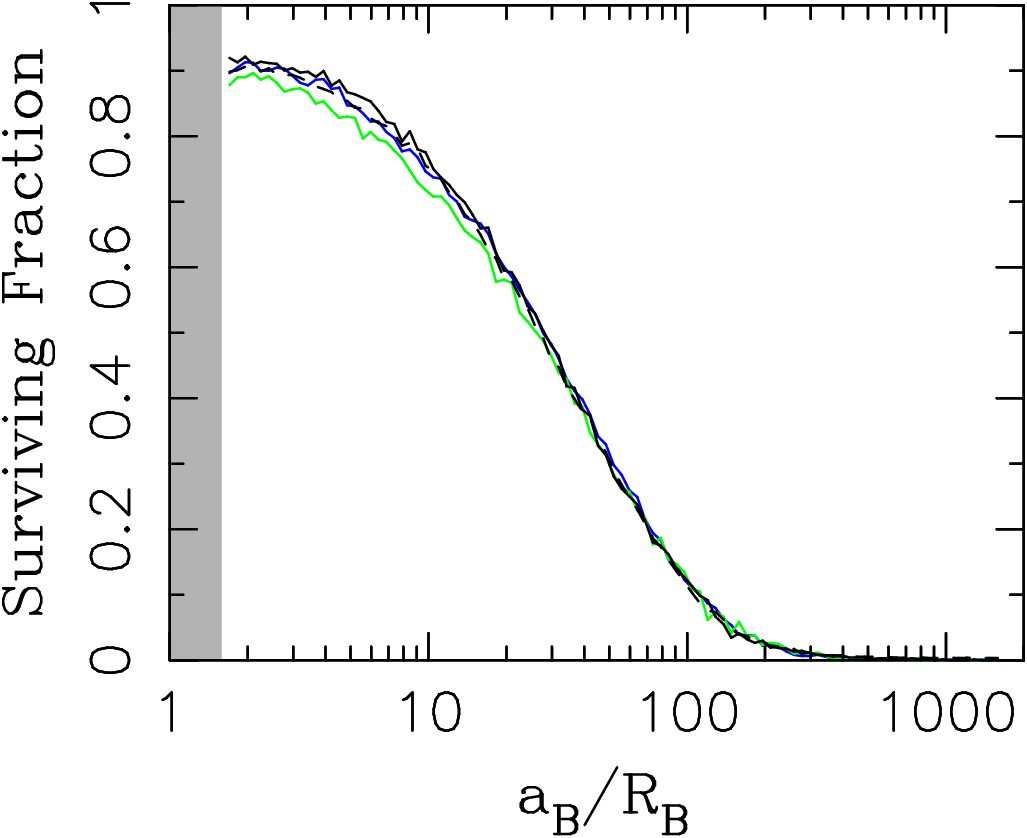}
\caption{The dynamical survival of equal-size binaries in different KBO populations:
HCs (blue), Plutinos (green), scattering (solid black) and detached (dashed black) objects.
The results obtained for different populations are very similar. The gray area denotes 
separations for which the binary components are in contact.}
\label{pops}
\end{figure}

\begin{figure}
\epsscale{0.8}
\plotone{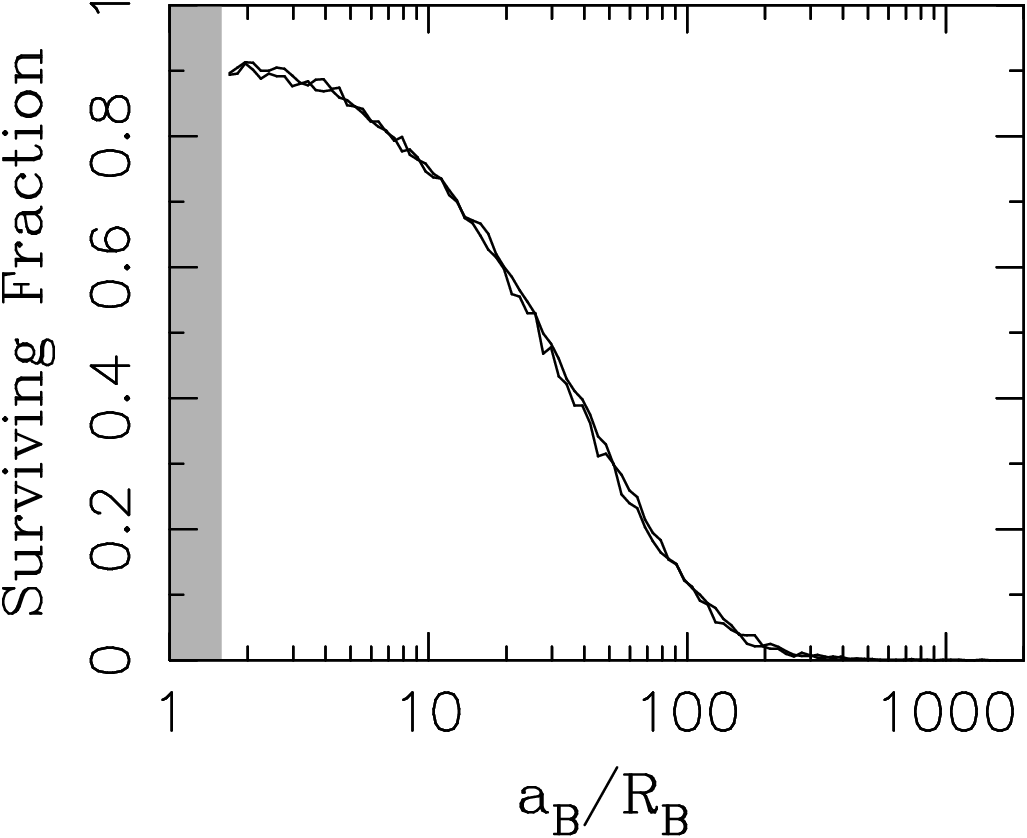}
\caption{The dynamical survival of equal-size binaries implanted into the HC population.
The solid curves show results for two different migration histories of Neptune (Table 1).
The results are very similar. The gray area denotes separations for which the binary 
components are in contact.}
\label{cases}
\end{figure}

\begin{figure}
\epsscale{0.8}
\plotone{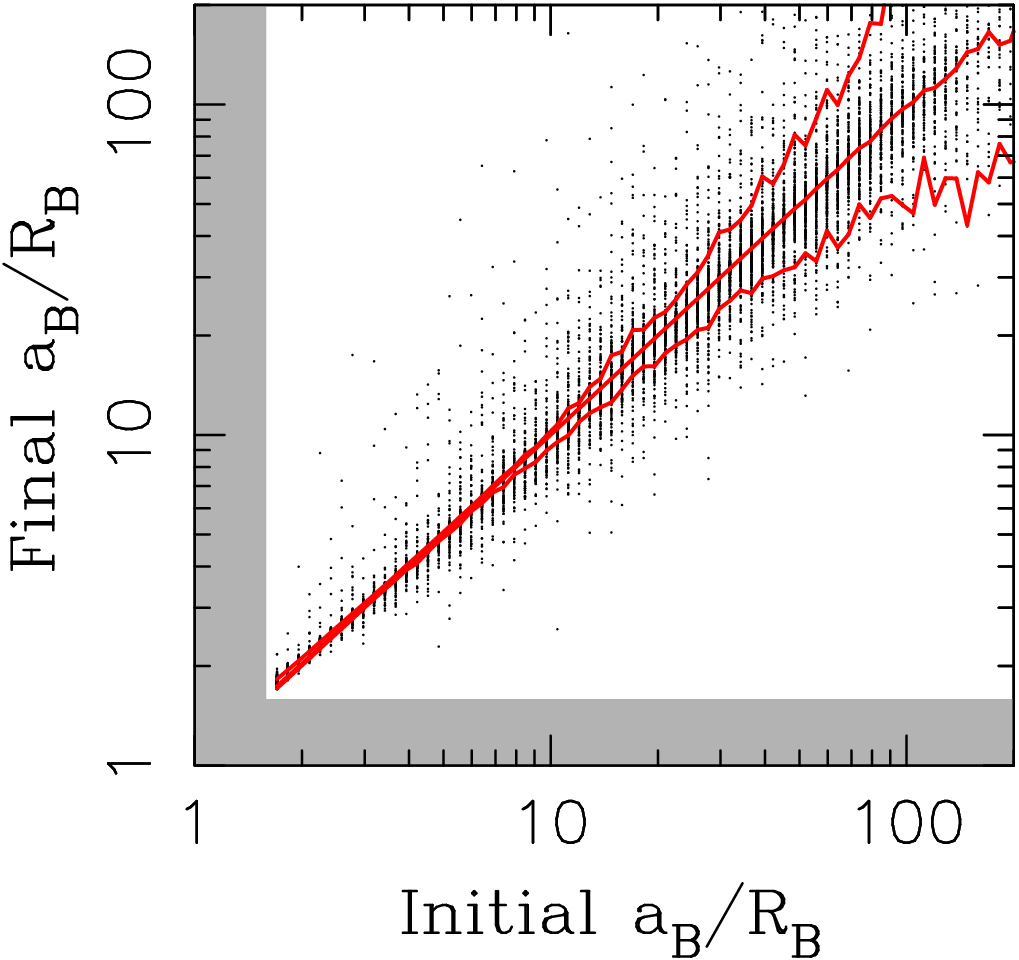}
\caption{The relationship between the initial and final separations of binary orbits in our simulations 
of planetary encounters. Only binaries surviving the whole sequence of encounters are shown here.
The black dots show individual cases. The red lines show the mean, and 5 and 95 percentiles. The gray 
area denotes separations for which the binary components are in contact. These results were obtained 
for the case2 migration parameters and HCs.}
\label{sema}
\end{figure}

\begin{figure}
\epsscale{0.7}
\plotone{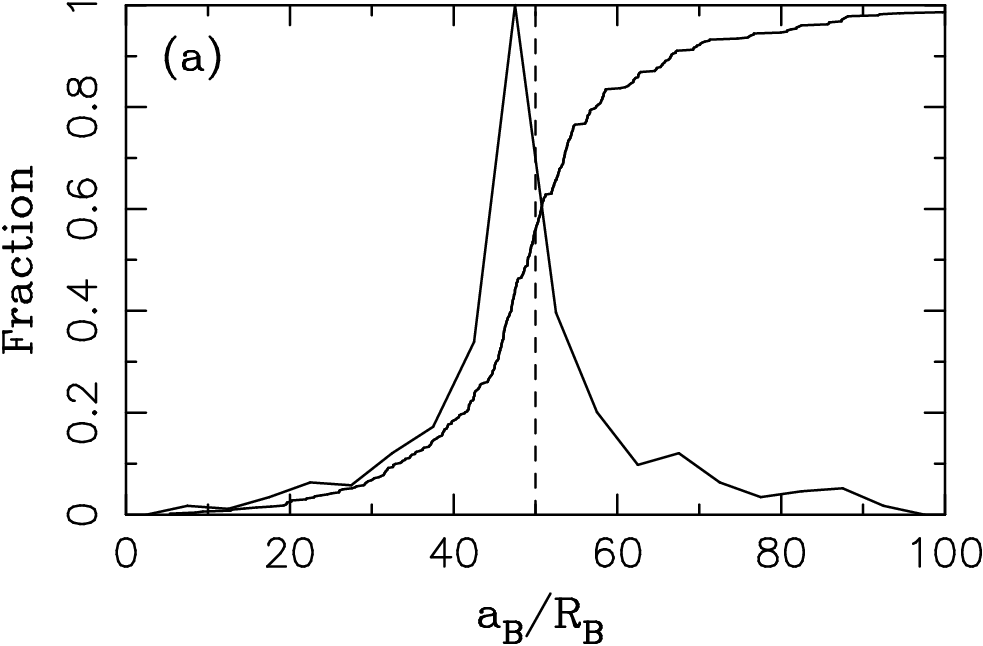}
\plotone{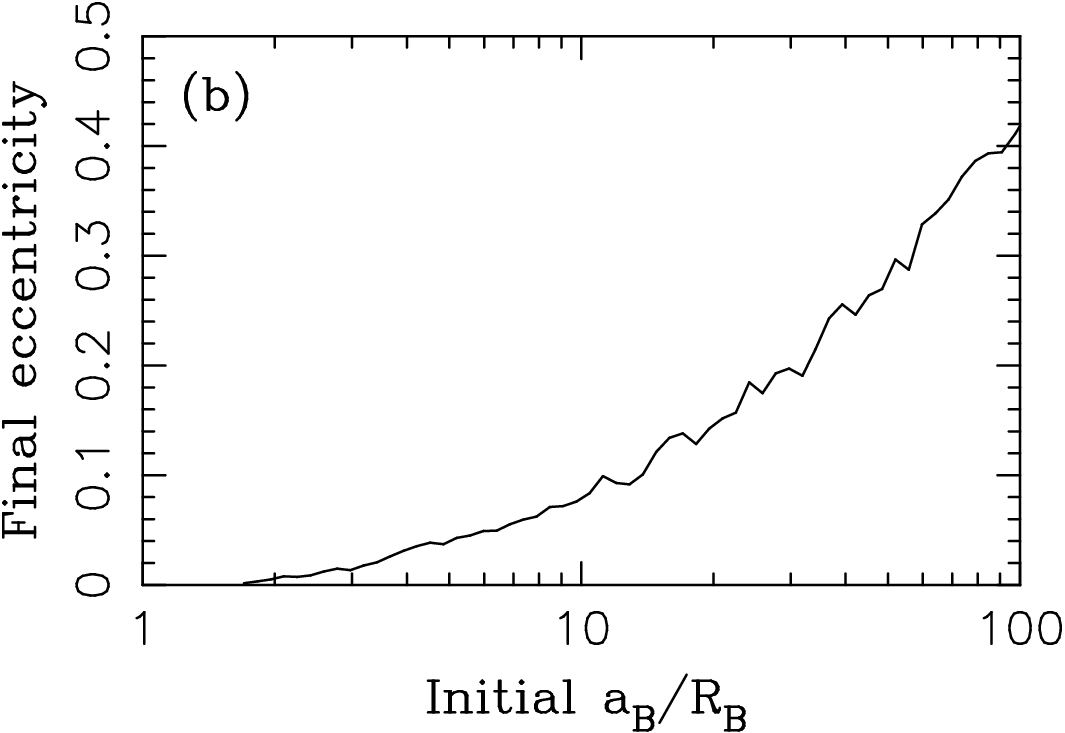}
\caption{The effect of planetary encounters on binary orbits. In panel (a), all binaries started with
$a_{\rm B}/R_{\rm B}=50$ (dashed line). The solid lines show the differential and cumulative 
distributions of $a_{\rm B}/R_{\rm B}$ of the final orbits. In panel (b), the final mean eccentricity of 
binary orbits ($e_{\rm B}=0$ initially) is shown as a function of initial $a_{\rm B}/R_{\rm B}$. 
The plotted eccentricity value was computed by averaging over all surviving binary orbits that started 
with the corresponding separation. The results were obtained for the case2 migration parameters and HCs.}
\label{ae}
\end{figure}

\begin{figure}
\epsscale{0.5}
\plotone{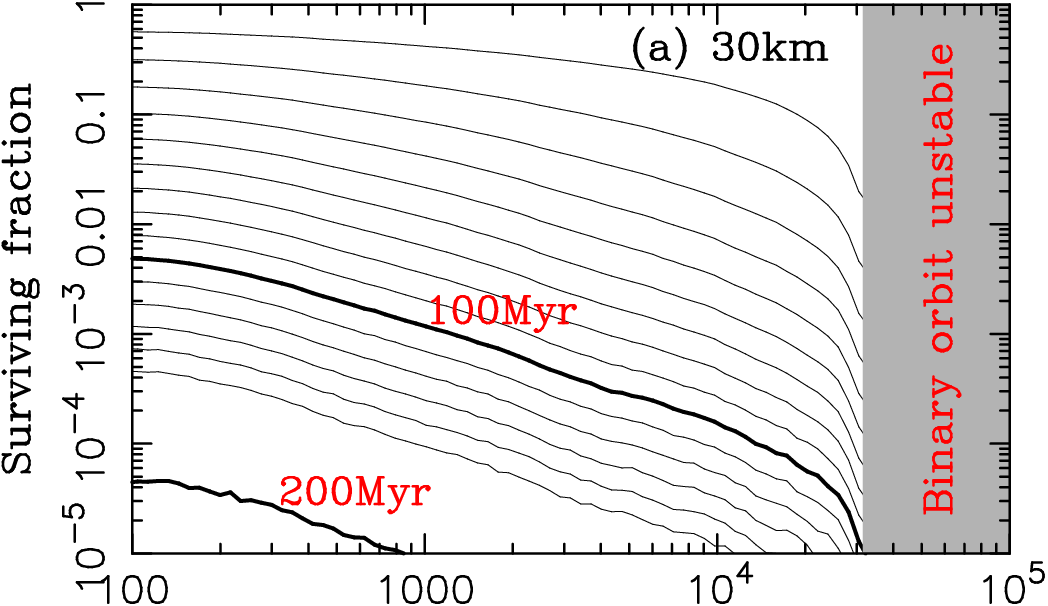}\\
\vspace*{3.mm}
\plotone{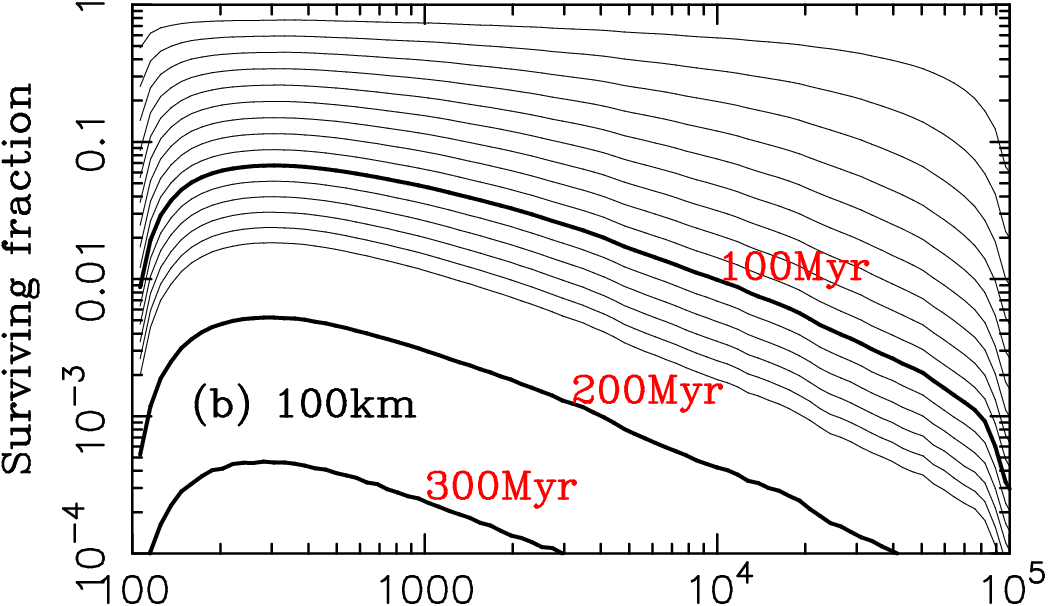}\\
\vspace*{3.mm}
\plotone{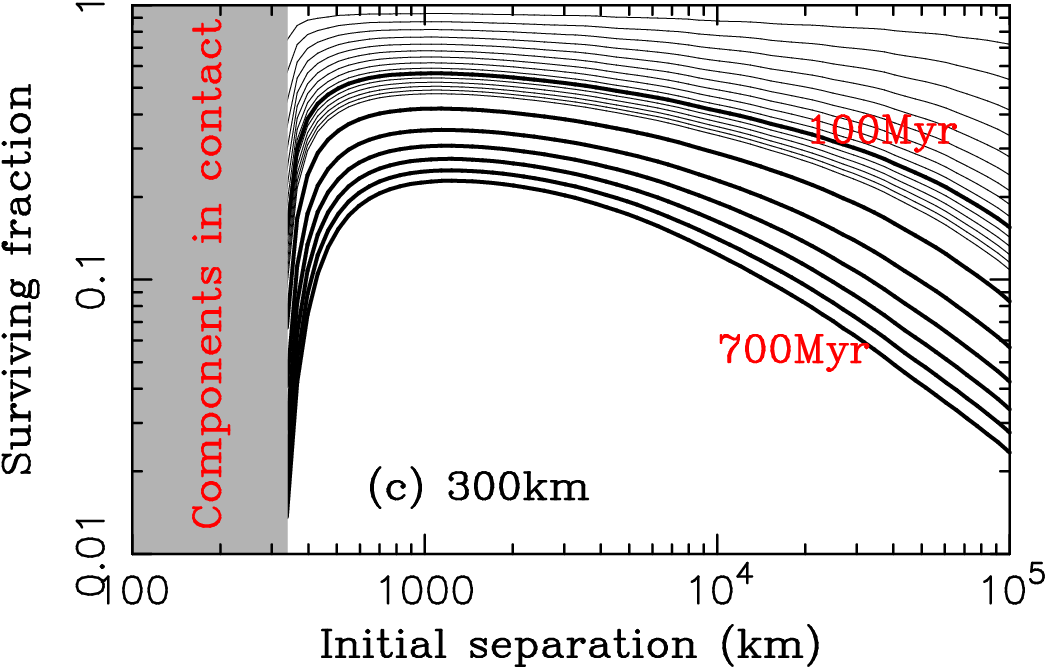}
\caption{The collisional survival of equal-size binaries in our nominal simulations of the 
massive planetesimal disk.
The panels illustrate the results for $R_1+R_2=30$ km (a), 100 km (b) and 300 km (c). The lines show the 
surviving fraction of binaries for different assumptions on the planetesimal disk lifetime, $t_{\rm disk}$
(thin lines are spaced by 10 Myr up to $t_{\rm disk}=150$ Myr, bold lines each 100 Myr).}
\label{col}
\end{figure}

\begin{figure}
\epsscale{0.45}
\plotone{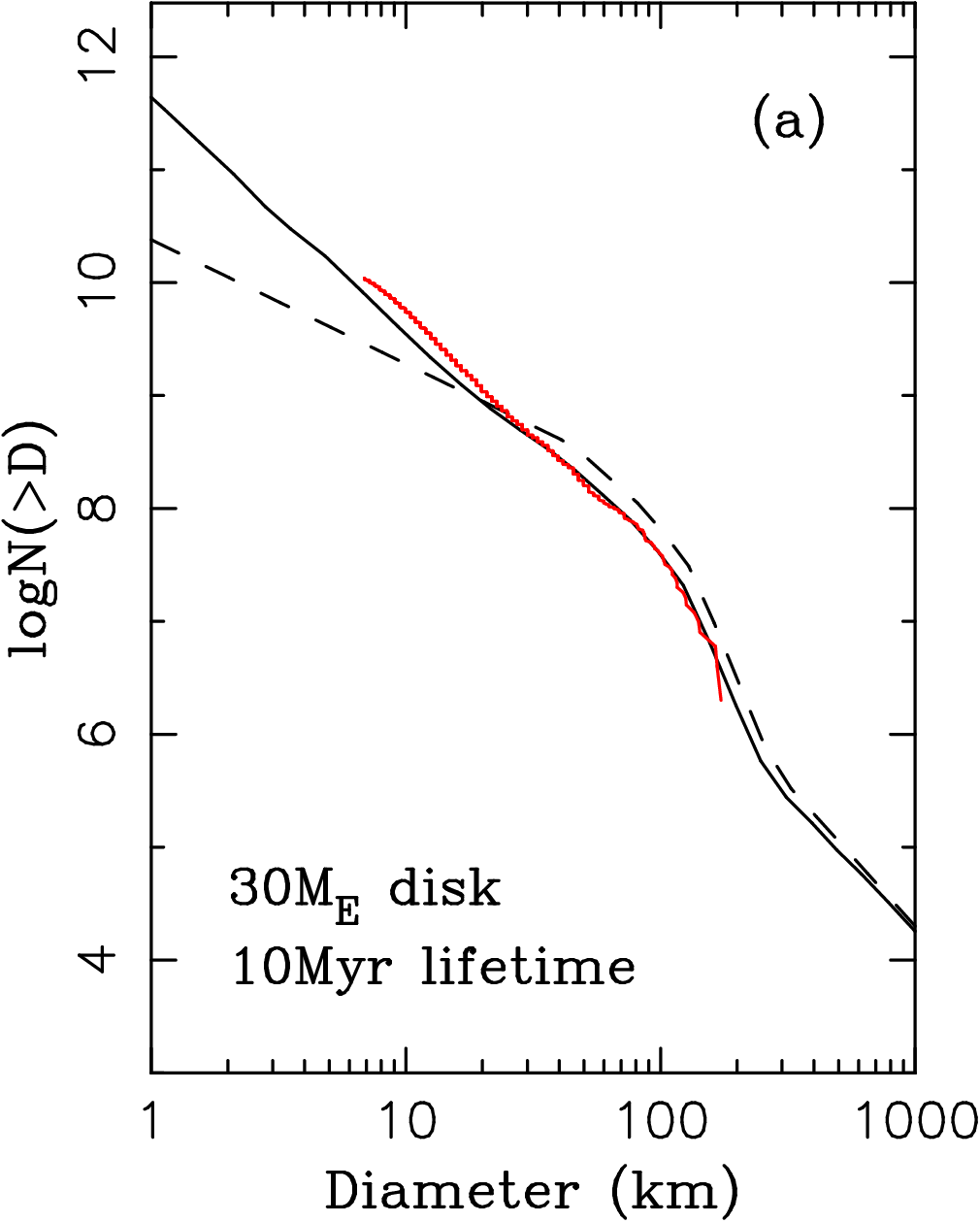}\hspace*{2.mm}
\plotone{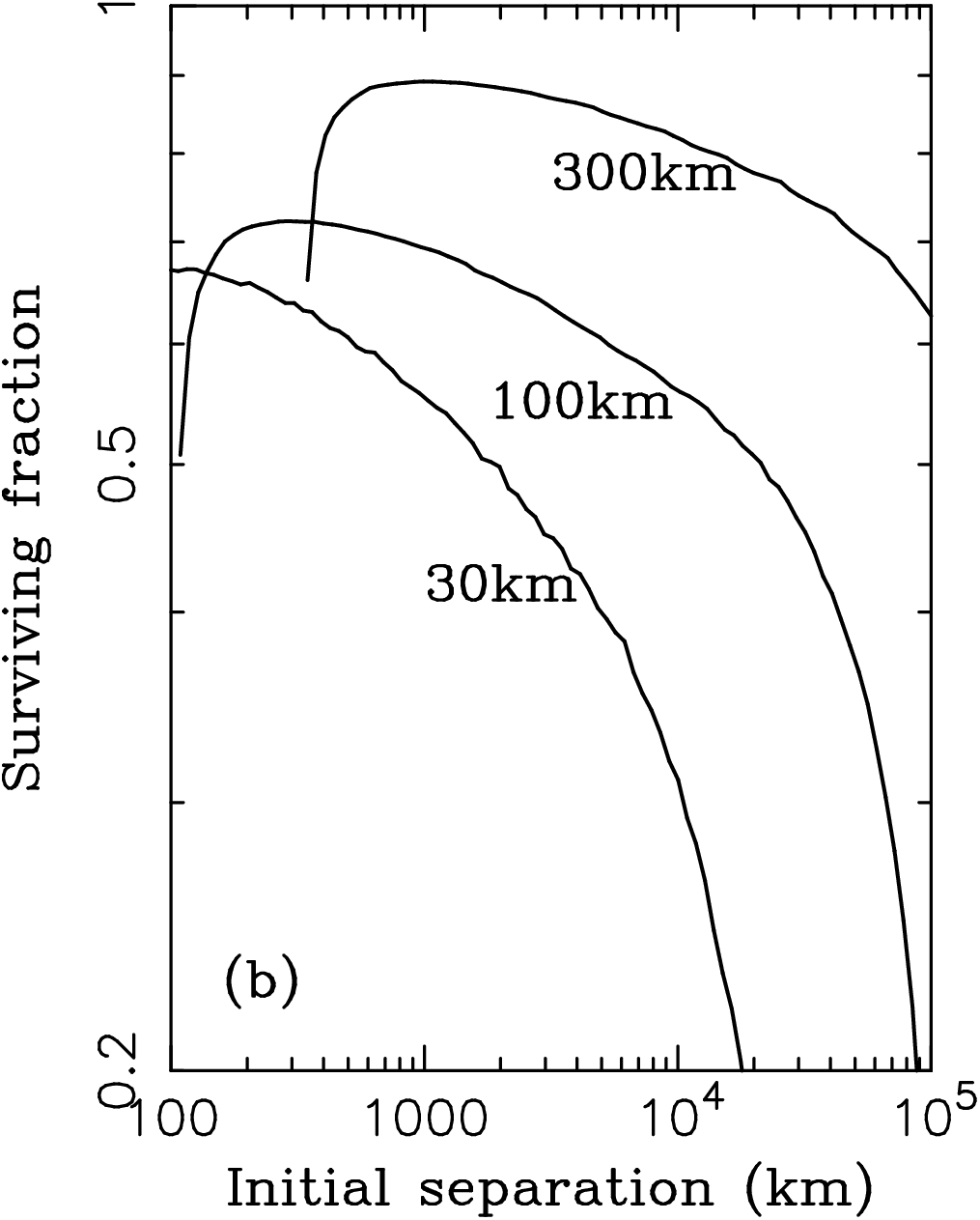}
\caption{Panel (a): collisional evolution of the outer planetesimal disk. The dashed line shows the 
initial size distribution. It corresponds to the initial disk mass $M_{\rm disk}=30$~$M_\oplus$. 
The solid line shows the size distribution after 10 Myr of collisional grinding, when $M_{\rm disk}=
21$~$M_\oplus$. Here we adopted the disruption law for ice from Benz \& Asphaug (1999) and $f_Q=0.1$.
The red line shows the size distribution of known Jupiter Trojans (incomplete for $D<10$ km) scaled up by 
their implantation efficiency (Nesvorn\'y et al. 2013). Panel (b): collisional survival of equal-size 
binaries in the disk. The lines show the results for $t_{\rm disk}=10$~Myr and $R_1+R_2=30$, 100 and 300 km.}
\label{si}
\end{figure}

\begin{figure}
\epsscale{0.45}
\plotone{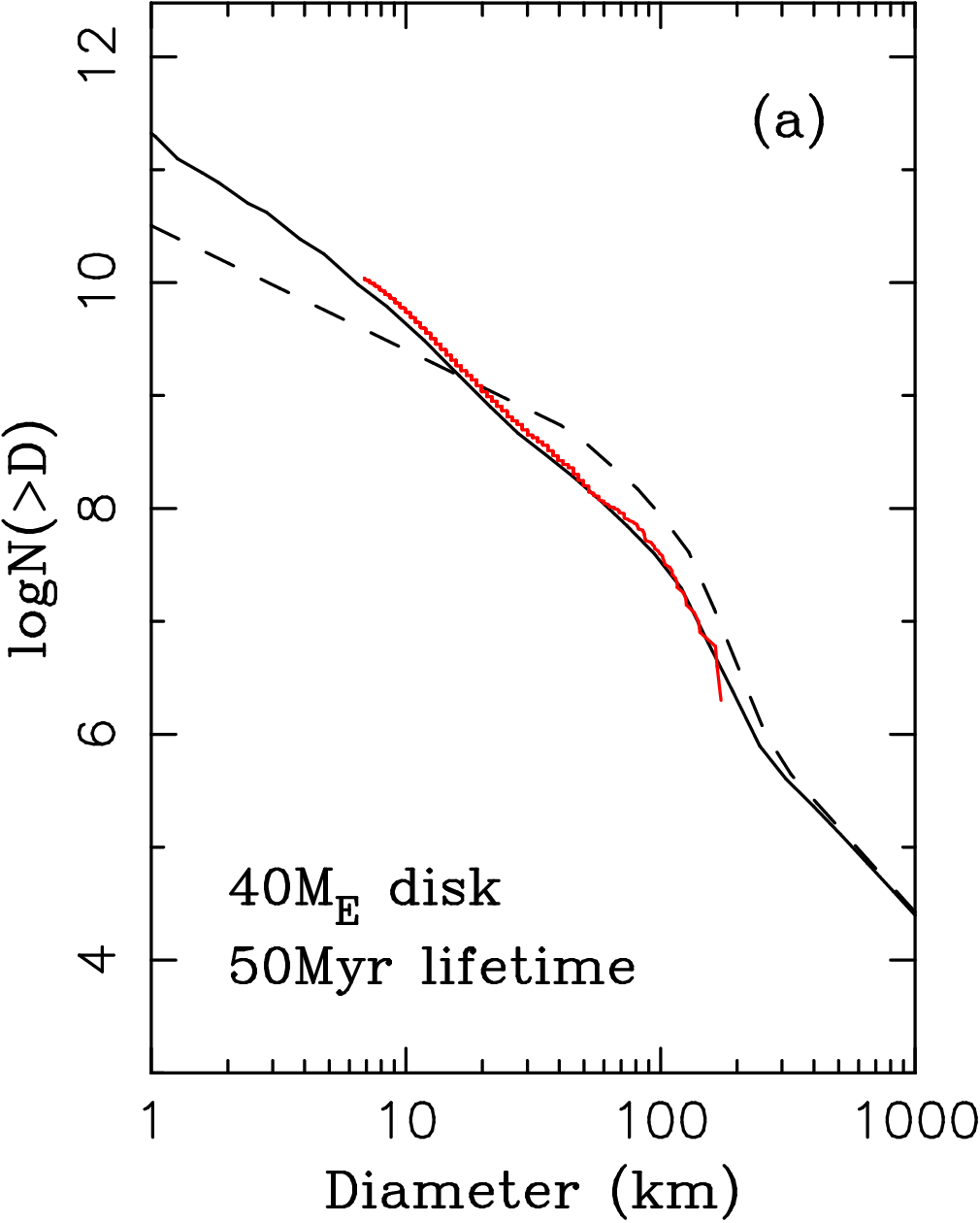}\hspace*{2.mm}
\plotone{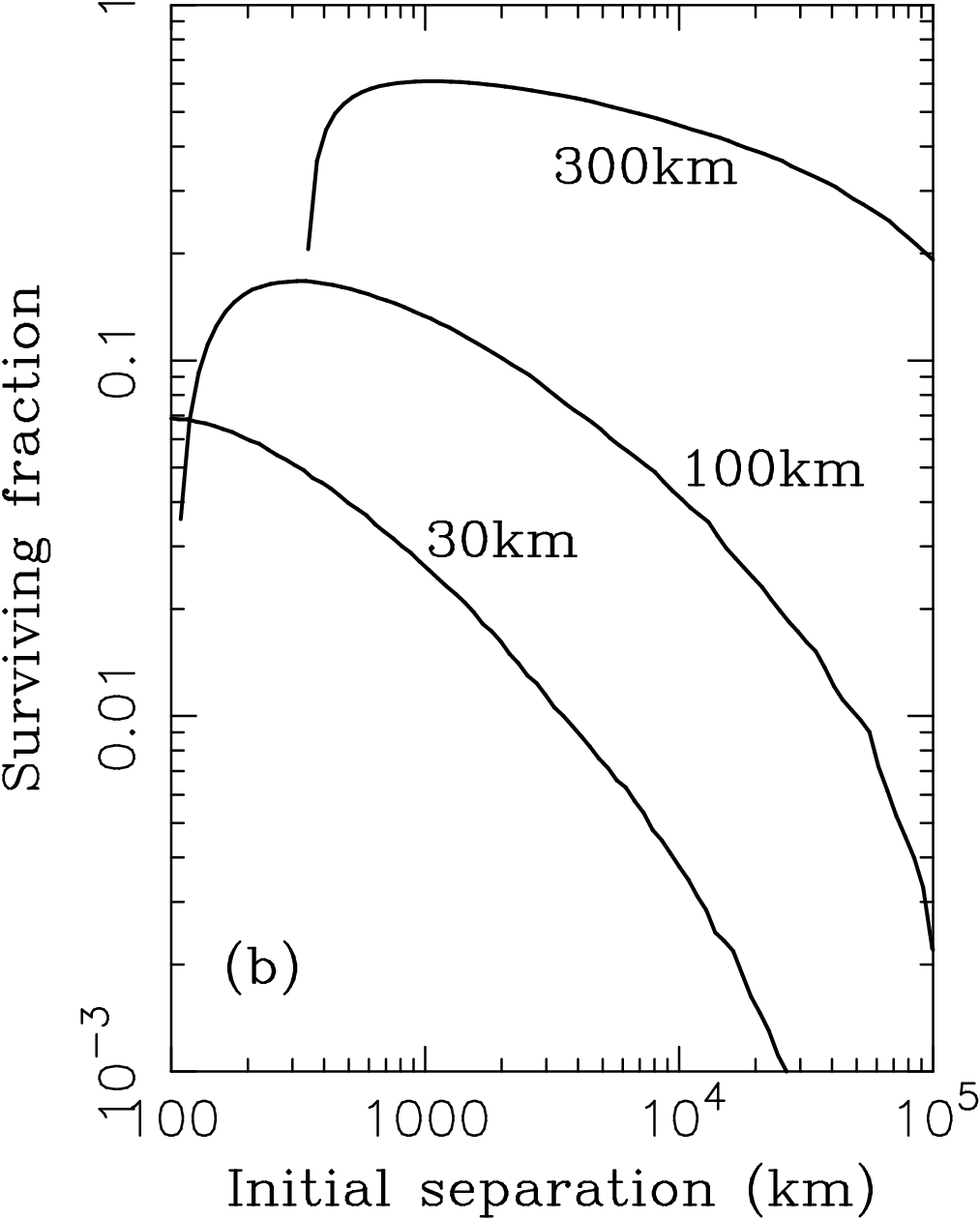}
\caption{The same as Fig. \ref{si} but for $M_{\rm disk}=40$ $M_\oplus$, $t_{\rm disk}=50$ Myr and 
$f_{\rm Q}=0.3$.}
\label{si2}
\end{figure}

\begin{figure}
\epsscale{0.6}
\plotone{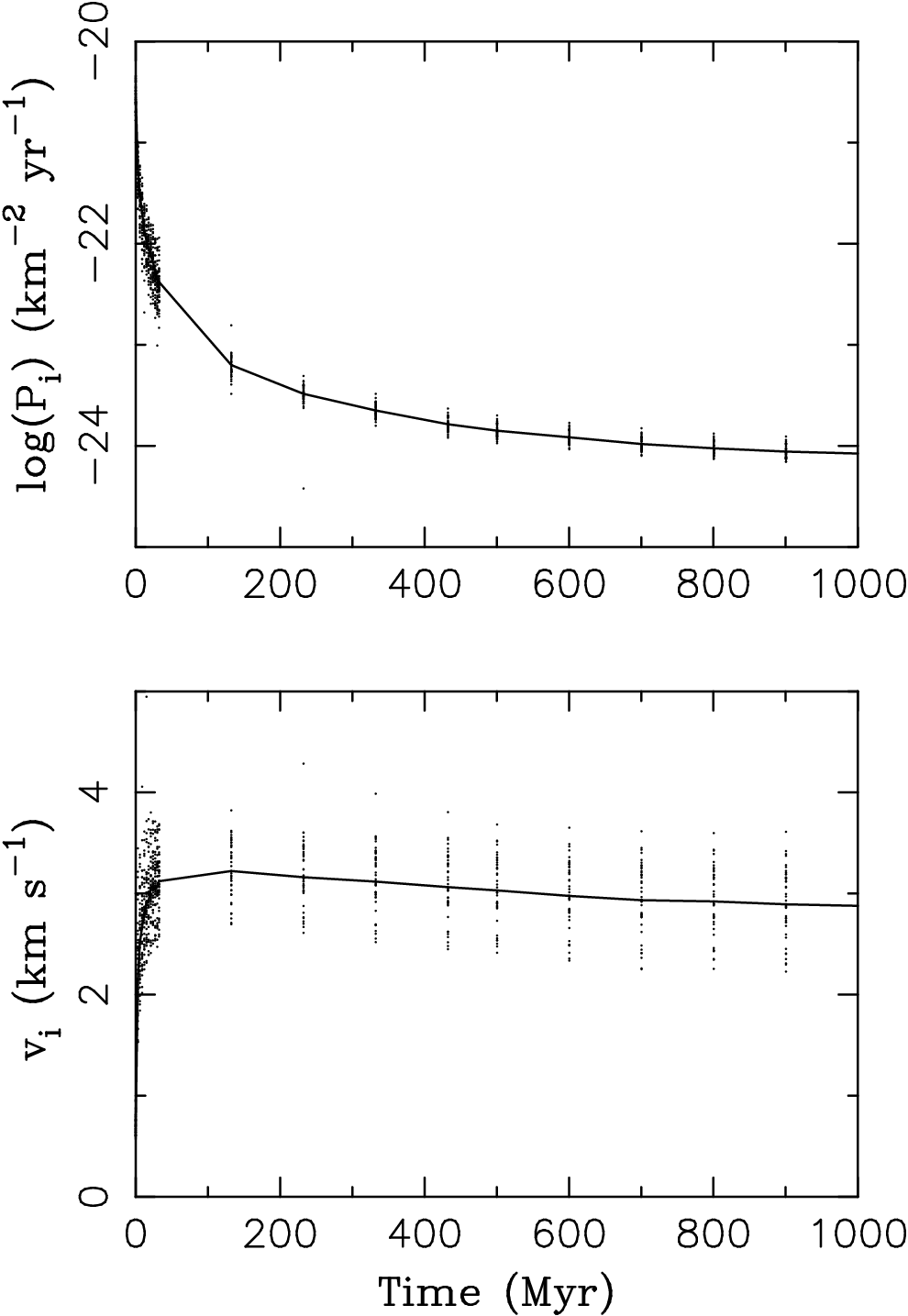}
\caption{The collision probability ($P_{\rm i}$) and impact speed ($v_{\rm i}$) of planetesimals. $P_{\rm i}(t)$ 
was normalized by the {\it initial} number of planetesimals. Its value at any given time therefore 
encapsulates both the changing collision probability and decaying population of planetesimals. Here
we assumed that the disk was dispersed by migrating Neptune immediately after $t_0$. Time $t=0$ 
in this plot therefore corresponds to the onset of Neptune's migration into the outer planetesimal disk.
At $t=0$, $P_{\rm i}=4\times10^{-21}$ km$^{-2}$ yr$^{-1}$ and $v_{\rm i}=0.6$ km, which is in the ballpark 
of the values considered for the simulations in Section 5.3. The normalized probability drops over 
time as the disk planetesimals are dispersed and eliminated from the solar system. Here, $P_{\rm i}(t)$ and 
$v_{\rm i}(t)$ are shown for the case2 migration parameters and HCs. The results for case1 and other
KBO categories are similar.}
\label{varcol}
\end{figure}

\begin{figure}
\epsscale{0.45}
\plotone{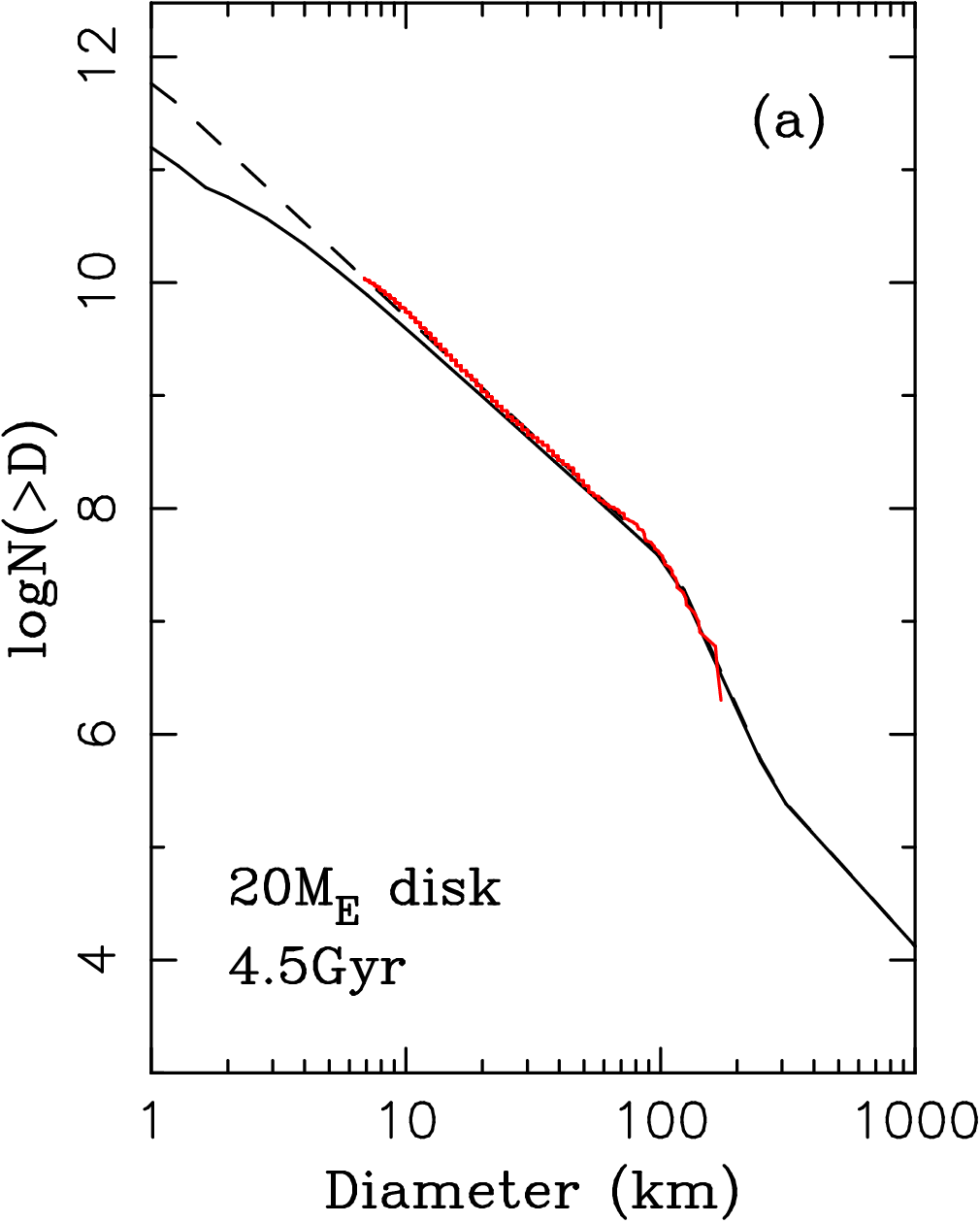}\hspace*{2.mm}
\epsscale{0.43}
\plotone{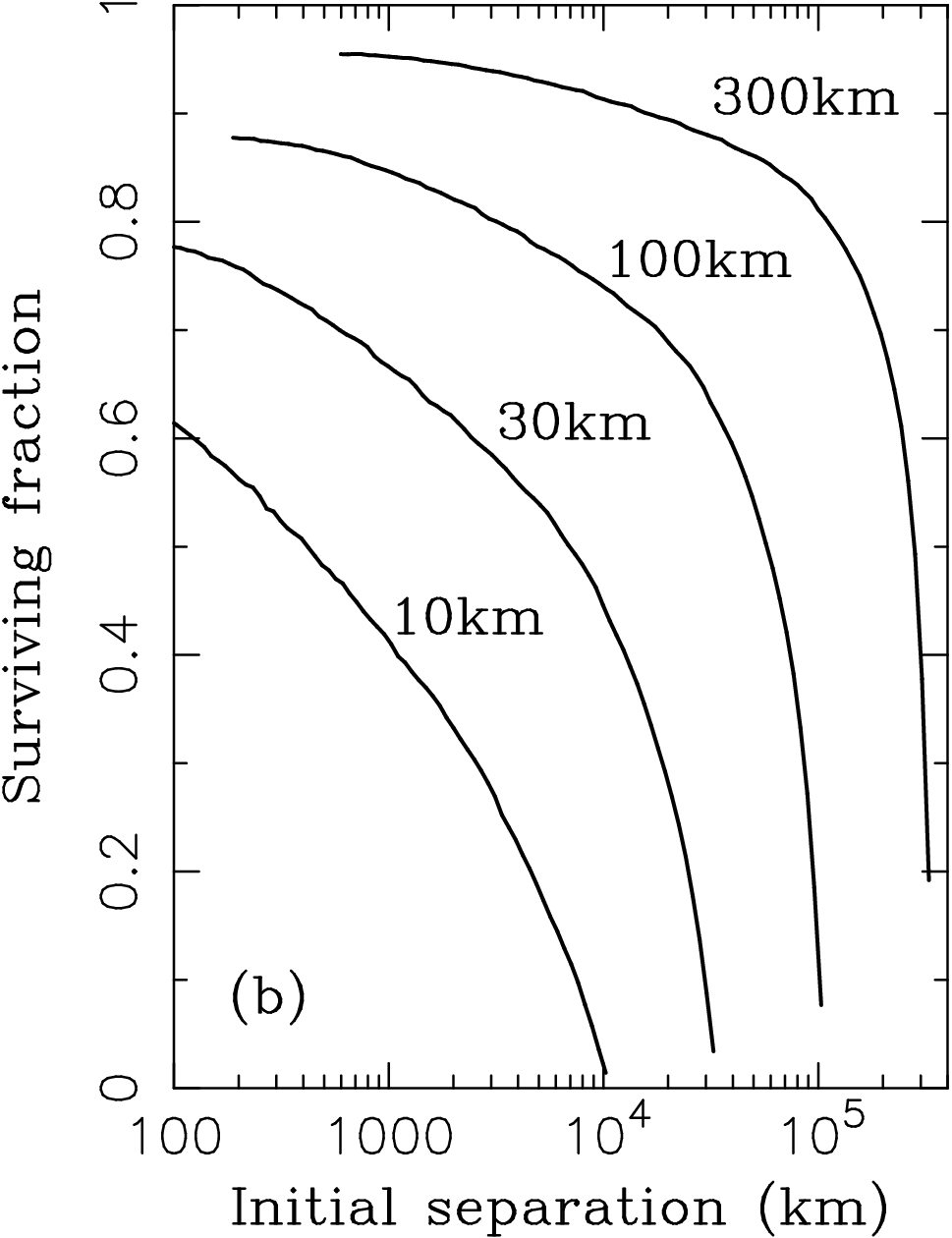}
\caption{The {\it Boulder} code results for $t_{\rm disk}=0$. In panel (a), we assumed that $M_{\rm disk}=20$ 
M$_\oplus$ at $t=0$ and the initial size distribution (solid line) similar to that of Jupiter 
Trojans (solid red line). The size distribution does not change much over time and the final distribution 
($t=4.5$ Gyr; solid black line) is similar to the initial one. In panel (b), the survival probability
is shown for the equal-size binaries with $R_1+R_2=10$, 30, 100 and 300 km (as labeled). These results
were obtained for $P_{\rm i}(t)$ and $v_{\rm i}(t)$ shown in Figure \ref{varcol}.}
\label{var1}
\end{figure}

\begin{figure}
\epsscale{0.6}
\plotone{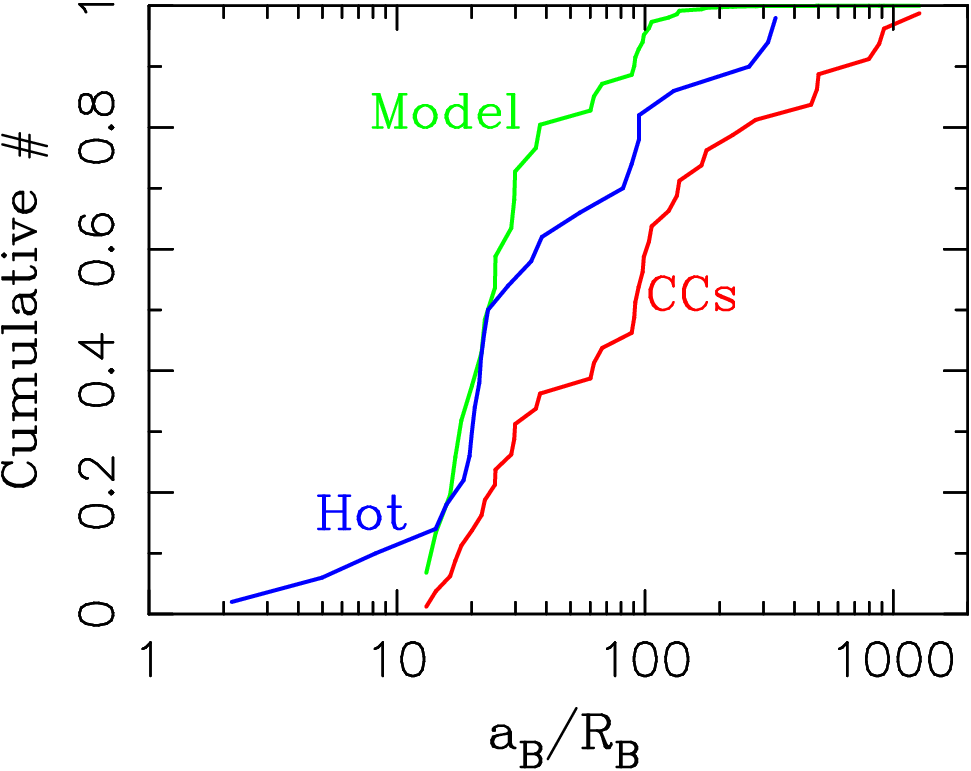}
\caption{The cumulative distribution of known equal-size ($R_2/R_1>0.5$) binaries in the dynamically 
cold (red line) and hot populations (blue line). The model distribution (green line) was obtained 
by applying the dynamical survival probability from Fig. \ref{scaled} to a template distribution
(see the main text).}
\label{model}
\end{figure}

We determined the dynamical and collisional survival of KBO binaries before and after their implantation 
into the dynamically hot population in the Kuiper belt. The main results are:
\begin{enumerate}
\item The binary survival is a strong function of the size and separation of binary components. The 
dynamical survival during scattering encounters with planets only depends on $a_{\rm B}/R_{\rm B}$,
where $a_{\rm B}$ is the binary semimajor axis and $R_{\rm B}^3=R_1^3+R_2^3$, with $R_1$ and $R_2$ 
being the radii of binary components. The tight binaries with $a_{\rm B}/R_{\rm B}<30$ are expected 
to survive in $>$50\% of cases, whereas the wide binaries with $a_{\rm B}/R_{\rm B}>100$ are expected 
to die in $>$90\% of cases (Fig. \ref{scaled}).
\item The existence of equal-size binaries in the dynamically hot population of the Kuiper belt
implies that the massive planetesimal disk below 30 au was short-lived ($t_{\rm disk}< 100$ Myr;
see also Nesvorn\'y et al. 2018). The disk could have started with a rounded size distribution
of planetesimals, as indicated by the streaming instability simulations (e.g., Simon et al. 2017),
and $M_{\rm disk}>20$ $M_\oplus$. It would subsequently collisionally evolve, within $\sim$10-50 Myr, 
to a size distribution similar to that of Jupiter Trojans and $M_{\rm disk}\simeq15$-20 $M_\oplus$ 
(Figs. \ref{si} and \ref{si2}). The size distribution and binary fraction are not expected 
to change much after the implantation of objects into the Kuiper belt (Fig. \ref{var1}).
\item The initial fraction of tight equal-size binaries in the massive disk should have been at least 
$\sim$18\% for $a_{\rm B}/R_{\rm B}<100$ to account for the present population of equal-size binaries
among the hot KBOs. The extremely wide equal-size binaries with $a_{\rm B}/R_{\rm B}>350$ have been  
removed during the implantation process ($<$0.003 survival probability). The existence of wide 
equal-size binaries with $100<a_{\rm B}/R_{\rm B}<350$ in the hot population is puzzling. They may 
survivors of a much larger original population of wide binaries in the massive disk, or, at least
in some cases, are interlopers from the dynamically cold population. 2002 VF130 with the
heliocentric orbit inclination $i=19.5^\circ$ and estimated $a_{\rm B}/R_{\rm B} \sim 310$ is an 
important constraint on the implantation process.    
\item All known satellites of the largest KBOs are expected to survive during the dynamical implantation 
of their primaries in the Kuiper belt. Most of them likely formed during the early stages when their
parent bodies were immersed in the massive planetesimal disk below 30 au and sustained intense bombardment. 
The low orbital eccentricities of Pluto's small moons may have been excited during encounters 
of the Pluto system to Neptune, or by small impacts during the massive disk lifetime. 
\item The expected fraction of contact binaries from the population of collapsed equal-size binaries 
is $\sim$10-30\% (this estimate assumes that all planetesimals formed as binaries), whereas the 
observational constraints indicate that the contact binary fraction among hot KBOs is 
$\sim$10-40\% (Sheppard \& Jewitt 2004, Thirouin \& Sheppard 2018).  
\end{enumerate}

\acknowledgements

We thank W. Grundy and K. Noll for the list of KBOs that were imaged by the HST. The work of D.N. was 
supported by the NASA Emerging Worlds program. The work of D.V. was supported by the Czech Science 
Foundation (grant 18-06083S). We thank J.-M. Petit and an anonymous reviewer for helful corrections 
of the submitted manuscript.

\clearpage
\begin{table}
\centering
{
\begin{tabular}{lrrrrrr}
\hline \hline
 number & temp. id.  & name  & KBO      & $R_1+R_2$   & $a_{\rm B}$  & $a_{\rm B}/R_{\rm B}$ \\  
        &           &       & class    & (km)       &  (km)       &                    \\  
\hline
(341520) & 2007 TY430 & Mors-Somnus         & Plu      & 100         & 21,000     &  335 \\
   --    & 2002 VF130 & --                  & HC       & 113         & 22,400     &  310 \\
   --    & 1998 WW31  & --                  & HC       & 135         & 22,620     &  263 \\
   --    & 2004 KH19  & --                  & HC       & 154         & 13,000     &  130 \\
(119067) & 2001 KP76  & --                  & HC       & 150         & 8,900      &   94 \\
   --    & 2004 PB108 & --                  & HC       & 187         & 10,400     &   81 \\
(47171)  & 1999 TC36  & Lempo               & Plu      & 202         & 7,411      &   50 \\
(82157)  & 2001 FM185 & --                  & Plu      & 129         & 3,130      &   38 \\ 
   --    & 2006 SF369 & --                  & SDO      & 142         & 3,120      &   35 \\    
\hline \hline
\end{tabular}
}
\caption{The wide, equal-size binaries in the hot population of the Kuiper belt 
(HC stands for Hot Classical, Plu for Plutino, SDO for a scattered disk object).
For (47171) Lempo (1999 TC36), which is a triple system, we list parameters of 
the outer component. All these binaries, except for 1998 WW31 (Veillet et al. 2002),
were imaged by the Hubble Space Telescope (Noll et al. 2008a,b; Grundy et al. 
2018).}
\end{table}

\clearpage
\begin{table}
\centering
{
\begin{tabular}{lrrr}
\hline \hline
                  & $\tau_1$ & $\tau_2$   & $N_{\rm Pluto}$  \\  
                  & (Myr)    & (Myr)      &            \\  
\hline
case1             & 30       & 100         & 4000       \\
case2             & 10       & 30          & 2000       \\
\hline \hline
\end{tabular}
}
\caption{A two stage 
migration of Neptune was adopted from Nesvorn\'y \& Vokrouhlick\'y (2016): $\tau_1$ and $\tau_2$ define 
the $e$-folding exponential migration timescales during these stages, and $N_{\rm Pluto}$ is the assumed 
number of Pluto-mass objects in the massive disk below 30 au. Neptune's migration is grainy with these objects
as needed to explain the observed proportion of resonant and non-resonant populations in the Kuiper 
belt.}
\end{table}

\clearpage
\begin{table}
\centering
{
\begin{tabular}{lrrr}
\hline \hline
                  & Charon   & Styx       & Hydra      \\  
\hline              
\multicolumn{4}{c}{ --case1--} \\
survive           & 0.94     & 0.87       & 0.82       \\
$e<0.1$           & 0.88     & 0.74       & 0.66       \\
$e<0.01$          & 0.82     & 0.66       & 0.53       \\
$0.003<e<0.012$   & --       & 0.04       & 0.04       \\
\hline
\multicolumn{4}{c}{ --case2--} \\
survive           & 0.91     & 0.82       & 0.74       \\
$e<0.1$           & 0.83     & 0.67       & 0.56       \\
$e<0.01$          & 0.75     & 0.56       & 0.41       \\
$0.003<e<0.012$   & --       & 0.05       & 0.05       \\
\hline \hline
\end{tabular}
}
\caption{The survival probability and orbital changes of Pluto's moons against dynamical perturbations during 
planetary encounters. The rows show the probability to satisfy different criteria. The results
for Nix and Kerberos are intermediate between those of Styx and Hydra. See Table 2 for definition of 
cases 1 and 2.}
\end{table}

\clearpage

\clearpage


\clearpage

\clearpage

\clearpage

\clearpage

\clearpage

\clearpage

\clearpage

\clearpage

\clearpage

\end{document}